# A microscopic view of the electromagnetic properties of sub–λ metallic surfaces

P. Lalanne[1], J.P. Hugonin[1], H.T. Liu[2], B. Wang[1]

[1]Laboratoire Charles Fabry de l'Institut d'Optique, CNRS, Univ Paris-Sud, Campus Polytechnique, RD 128, 91127 Palaiseau cedex, France
[2]Key Laboratory of Opto-electronic Information Science and Technology, Ministry of Education, Institute of Modern Optics, Nankai University, Tianjin 300071, P. R. China.

**Abstract**
We review the properties of the surface waves that are scattered by two–dimensional sub–λ indentations on metallic surfaces. We show that two distinct waves are involved, a surface plasmon polariton (SPP) and a quasi–cylindrical wave (quasi-CW). We discuss the main characteristics of these waves, their damping characteristic lengths and their relative excitation weights as a function of the separation distance from the indentation and as a function of the metal conductivity. In particular we show that derive a closed–form expression for the quasi-CW, which clarifies its physical origin and its main properties. We further present an intuitive microscopic model, which explains how the elementary SPPs and quasi-CWs exchange their energies by multiple–scattering to build up a rich variety of near– and far–field optical effects.

Key–words : sub–λ metallic surfaces; surface–plasmon–polaritons; quasi–cylindrical waves; plasmonics; extraordinary optical transmission; diffraction; multiple scattering

## *1. Introduction*

For more than a century, the optical response of metal–dielectric interfaces has constituted a topic of many experimental and theoretical investigations. This response encompasses a rich variety of phenomena, which manifest themselves as huge variations in the reflection, absorption, transmission of incident fields [1] or in the emission of molecules located nearby [2]. As in many other scientific areas, periodic structures have played a major role in the elaboration of the underlying theoretical concepts and of the associated applications. Doubtlessly, one of the first milestone achievements has been the interpretation of Wood's anomalies (discovered at the beginning of the 20[th] century [3]) as resulting from a Fano–type excitation [4] of surface modes on relief gratings [5]. As for the simplest case of flat interfaces [6], the surface plasmon polariton (SPP) corresponds to a collective resonance of localized electromagnetic fields and of charge density oscillations in the metal. They are electromagnetic modes (i.e. solutions of Maxwell's equations in the absence of any source), which are mathematically described as the complex poles of the scattering operators associated to the corrugated surface [5]. The surface modes may be either bounded to the surface or they may radiate into one or more radiating space harmonics, explaining the anomalies initially observed. The pole interpretation represents a macroscopic analysis, in the sense that the surface resonance is described as a collective response of the *periodic* surface [5]. In particular, the macroscopic character reflects into the mathematical and numerical tools that are developed for calculating the poles; the latter are associated to scattering operators that link macroscopic (observable) quantities, such as the amplitudes scattered into the different orders [1,7], which are defined for the entire *periodic* structure.

These initial works have been mostly carried out for analyzing reflection gratings with continuous profiles, often for reducing the negative effect of the efficiency anomalies for the performance of spectroscopic instruments [1,8]. With the recent advances in nanotechnology and in relation with the observation of the extraordinary optical transmission (EOT) through arrays of tiny holes [9], the last decade has seen a renewed interest in exploiting the optical response of metallic materials with sub−λ openings. Initial motivations were the trapping of light in small volumes [10], various applications of the EOT [11−12], and waveguiding with lateral mode sizes well below the diffraction limit for ultra−compact SPP−based photonic circuits [13−15]. This has initiated the creation of a variety of optical components for manipulating SPPs at wavelength scales.

Our current understanding and design recipes all rely on models based on the scattering of individual SPPs, which are assumed to be first generated by some illuminated apertures, then to propagate on the metal surface and to interact with nearby indentations before being recovered as freely propagating light or detected. This model is readily illustrated by considering the Young's double−slit experiment shown in the inset of Fig. 1. The far−field interference patterns observed for the transverse−magnetic (TM) polarization, when only a single slit of the doublet is illuminated by a focused Gaussian beam (see the inset), is shown in Fig. 1. The calculation has been performed for several wavelengths, assuming the gold permittivity−dependence given by the tabulated data in [16]. Although a single slit is illuminated, the far field exhibits interferences that are nothing else than the classical Young's interferences with a fringe spacing inversely proportional to the slit separation distance. At visible frequencies (top curves), a simple way to understand the fringe formation is to consider that the incident Gaussian beam transmits light into the illuminated slit and launches a SPP that propagates to the second slit, where it excites the second slit. The later acts as a secondary source that produces fringes by interference with the light scattered by the first slit in the far field. Indeed the SPP multiple−scattering model [17−24] is conceptually powerful and represents a microscopic vision that helps designing compact plasmonic devices [25−33]. The question arises how do we model the slit doublet experiment in practice, how do we calculate the far field response for instance?

There are many numerical tools (some are purely numerical like finite−element and finite−difference methods, some are more physically oriented like modal− or multipole−expansions methods …), but to our knowledge, none of them *explicitly* consider the SPPs: the main physical ingredients of our understanding of the phenomena (the SPP launching, propagation and scattering) are only implicitly taken into account in the modelisation by matching the continuous electromagnetic fields quantities at the interface. Clearly, *a gap exists between the way we understand the phenomena and the way we model it.* The gap even further increases as one moves from slit doublets to slit ensembles. For periodic slit arrays for instance, our present numerical tools, such as modal expansions techniques, and our concepts, such as the Fano–resonance phenomenological interpretation, are so far from the microscopic picture that, except in some rare works [34], one usually abandons the SPP multiple–scattering picture even at a conceptual level. As a consequence, our understanding of many optical important properties of nanostructured surfaces, such as the efficient transmission through metallic gratings composed of tiny slits (actually, the fraction of the transmitted photons far exceeds the fraction of the photons that are impinging onto the slits and virtually 100% of the incident photons pass through the grating for some resonance frequencies) is considerably reduced. This might explain why the past decennia has seen a strong debate [35] on this apparently simply phenomenon, known for a long time and even used by Hertz himself in the 1800's for testing the newly discovered radio–waves.

Indeed, numerical and conceptualization tools have to be intimately coupled. One of the objectives of this report is to try to reduce the present gap. Ideally, one would like to derive the analogue of the tight–binding model in solid–state physics for surface waves on patterned metallic surfaces, starting from the limit case of elementary scattering events of the individual microscopic sub–$\lambda$ indentations (the atomic orbitals), to move to macroscopic properties of the surface (the band structure, the far field radiation diagram …).

To fully exploit the potential of nanoscale metallic devices, it is crucial to achieve a precise vision of the electromagnetic interaction responsible for multiple scattering processes on corrugated surfaces. SPPs are only partly responsible for this interaction, and many optical phenomena, which are observed with metallic nanostructures at visible frequencies, mainly because of SPPs, can be reproduced in the THz and microwave domains by scaling the geometrical parameters without any SPP enrolments [36–37]. This statement can be easily realized by considering again the Young's double slit experiment. Indeed at visible and near-infrared frequencies (top curves), the SPP–excitation at the exit of the illuminated slit and its scattering by the other slit is responsible for the interference [19–20]. But at longer wavelengths, the SPP excitation efficiency vanishes [18], and SPPs cannot be considered to explain the fringe pattern. In fact at infrared frequencies, the interaction between the slits is mediated by another wave, called quasi–cylindrical wave (quasi-CW) hereafter. The nature of the quasi-CWs is very different from that of the SPPs [37], but since quasi-CWs are launched with an initial phase and intensity similar to those of SPPs and since it propagates with almost the same phase velocity, the fringe pattern remains essentially unchanged as one strides the entire spectrum. In general, SPPs and quasi-CWs are equally important and the dual–wave picture is needed to understand the physics or to carefully engineer plasmonic devices involving multiple scattering processes on corrugated surfaces.

In relation with this microscopic description, it is the purpose of this review to examine the concepts, to elucidate the underlying physics and to discuss recent results and current problems, with particular emphasis placed on work carried out in the last decade. In Section 2, we examine how much SPP is launched on metallic surfaces by sub–$\lambda$ two–dimensional (2D) individual indentations under illumination by an incident plane wave. We further present a general formalism that allows us to calculate the amount of SPPs generated on the flat–interface sections in between two nearby indentations of any array. Section 3 is devoted to a precise definition of the quasi-CW and to a study of its main properties. This section that contains original materials is motivated by the fact that the quasi-CW has been only recently discussed in the literature and that different visions are presently available. In particular, by analytically solving the problem of the emission of a line source on a metal surface (Appendix A), we clearly define what is meant by "quasi" and show that the quasi-CW can be seen as the electromagnetic field associated to the limit case of a plane wave impinging at grazing incidence. Section 4 examines the scaling laws associated to the generation strengths of SPPs and quasi-CWs by sub–λ indentations, as one varies the wavelength of the excitation field and the geometrical parameters accordingly. The net benefit is a comprehensive view of the dual-wave vision over the entire electromagnetic spectrum. In Section 5, we study the scattering of SPPs and quasi-CWs by nano−indentations, emphasizing cross−conversion inelastic processes that convert SPPs into quasi-CWs and vice versa. Under the assumption that the indentations have sub−λ dimensions, we define scattering coefficients for the SPPs, quasi-CWs, and for a combination of theses waves. This allows us to propose a microscopic description of the electromagnetic properties of sub−λ metallic surfaces, which generalizes the simplistic classical model based on the scattering of SPPs. Section 6 concludes the review.

The numerical data presented throughout the review are all obtained with a fully−vectorial frequency−domain modal method relying on Fourier expansion techniques. The aperiodic Fourier−Modal method [38−40] (a-FMM) is a generalization of the Rigorous Coupled Wave Analysis [41], which relies on an analytical integration of Maxwell's equations along one direction and on a supercell approach with Perfectly−Matched−Layers in the two other transverse directions. Hereafter, the metal is considered as a real metal with a finite conductivity. Gold will be used to illustrate our discussion and its frequency-dependent permittivity $\varepsilon_m=(n_m)^2$ is taken from [16]. Let us add that the analysis remains valid for other noble metals. Hereafter, we drop the vacuum impedance, so that $Z_0=(\mu_0/\varepsilon_0)^{1/2}=1$. This amounts to normalize the electromagnetic fields, $\mathbf{E}\leftrightarrow\mathbf{E}/(Z_0)^{1/2}$ and $\mathbf{H}\leftrightarrow\mathbf{H}\times(Z_0)^{1/2}$.

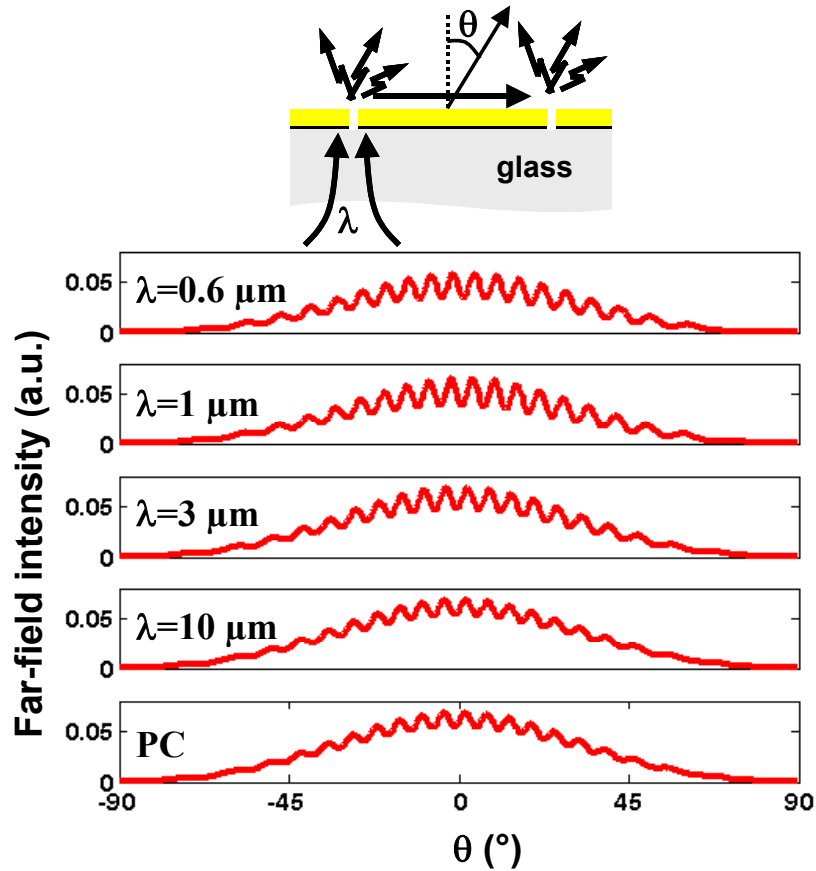

**Fig. 1.** Far-field fringe pattern of a slit doublet experiment when solely a single slit is illuminated by a focused Gaussian beam. The calculation has been performed with the a-FMM for several wavelengths ranging from the visible (top) to the infrared (bottom) by scaling all the geometrical parameters with the wavelength. The inset shows the doublet geometry in a gold film deposited on a glass substrate covered by a titanium adlayer. The waist of the normally incident Gaussian beam is $2\lambda$, and transverse-magnetic (magnetic-vector parallel to the slit) polarization is considered. No fringe is obtained for the other polarization. The slit widths are $0.4\lambda$, the slit separation distance is $10\lambda$, the gold thickness is $\lambda/4$ and the adlayer thickness is $\lambda/160$. Except for the bottom curve where a perfect conductor (PC) is considered, gold is considered as a real metal with a finite conductivity and with a frequency−dependent permittivity $\varepsilon_m$ taken from [16]. SPPs are mainly responsible for the fringe existence at $\lambda$=600 nm, partly involved in the multiple-scattering process at $\lambda$=1 μm, and very weakly excited in the infrared. They cannot be considered for explaining the fringe pattern predicted for the PC case. Experimental results supporting the numerical predictions at visible frequencies can be found in [19–20].

## *2. Generation of surface−plasmon−polaritons*

Under illumination by an incident light, a perfectly flat surface cannot launch SPPs on the surface. Surface roughness is needed to excite collective electron states [6,42]. 2D sub−$\lambda$ objects (such as grooves or ridges) on a metal film may launch a pair of counter−propagating SPPs on both sides of the object. This has been known for a long time and one commonly uses scratches to couple or decouple bounded SPPs from metallic samples [43−48].

However, the quantitative evaluation of how much SPP is launched has been investigated only recently. In this Section, we first present a theoretical formalism [18] for rigorously calculating the amount of SPPs that are launched by 2D illuminated indentations and then summarize known results for slits, grooves and ridges.

### **2.1. General fully−rigorous formalism**

In this Section, we present a theoretical formalism for the calculation of the excitation coefficients of SPP modes on metallic surfaces patterned by an arbitrary array of 2D indentations. The formalism is general. It remains perfectly rigorous over the whole spectrum, as it applies to tightly-bounded SPP modes at visible frequencies or to weakly confined SPPs at longer wavelengths in the infrared. It may be used for individual scatterers or for arrays of them. For arrays of finite lengths, not only it predicts the amount of SPP that is excited on both sides of the arrays (like for individual scatterers) but also it can be applied to calculate the SPPs that are excited inside the array in between the sub−$\lambda$ indentations on the flat parts of the interfaces. Over the past few years, it has been successfully applied to various geometries, including grooves and slits [18,49], metallic or dielectric ridges [50], or more sophisticated patterns such as hole chains [34].

To illustrate our purpose, let us consider the scattering by a sub−$\lambda$ slit in a gold film, see Fig. 2a. The slit is illuminated by its fundamental guided mode ($\lambda$=700 nm) with a polarization perpendicular to the slit (TM polarization) and with a unit amplitude at $x=y=0$. Figure 2b shows the magnetic field $H_z(x,y)$ calculated with the a-FMM. As expected, the scattered field of the sub−$\lambda$ 2D indentation essentially looks like a cylindrical wave. This holds only above the surface in the space region where only propagative waves are present. Indeed a careful inspection of the scattered pattern in the vicinity of the surface reveals a slight field enhancement with a marked expansion of several wavelengths away from the surface. This observation is consistent with the involvement of SPPs at the slit aperture. However, the respective roles of the evanescent fields, the propagative fields and of the SPP modes deserve to be clarified.

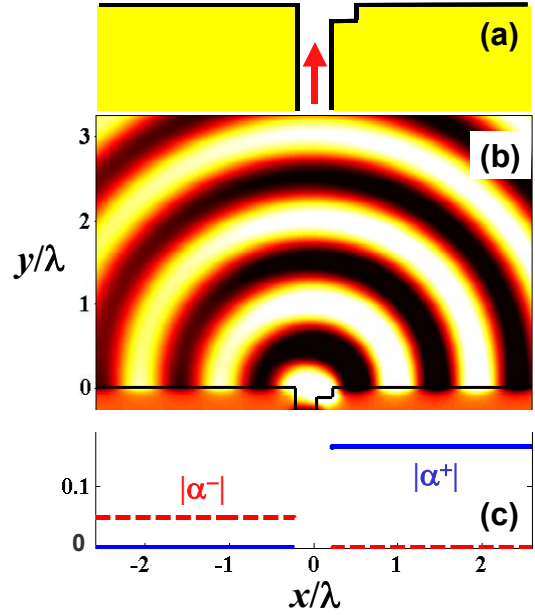

**Fig. 2.** SPP-field scattered by a sub–λ indented slit in a gold substrate illuminated by its fundamental mode with a unitary magnetic field at $x=y=0$. (a) Schematic of the single slit configuration. (b) Scattered magnetic field, Re($H_z$). (c) SPP generation strengths $|\alpha^+|$ and $|\alpha^-|$ obtained from the near-fields patterns shown in (b) by calculating the overlap integrals of Eqs. (4) and (5). The calculation are performed for a slit width $w=0.25\lambda$ ($\lambda=800$ nm) and an indentation of width 0.187λ and height 0.125λ.

To explore this question, we refer to the completeness theorem of the normal mode set used in waveguides theory [51]. That theorem, which provides a useful electromagnetic representation of light propagation in systems composed of a set of waveguide sections, stipulates that any transverse field pattern of any section can be decomposed as a linear combination of forward– and backward–travelling bounded and radiative modes of the considered section. For our slit geometry, two sections corresponding to a flat interface can be considered outside the slit aperture. From the completeness theorem, it results that the electromagnetic fields $\psi=[H_z,E_x,E_y]$ of Fig. 2b can be expanded into the complete set of normal modes of a flat interface. More specifically, everywhere outside the slit aperture, one may write

$$\psi(x,y) = [\alpha^+ \Psi_{SP}^+(x,y) + \alpha^- \Psi_{SP}^-(x,y)] + \Sigma_\sigma a_\sigma H_\sigma(x,y), \qquad (1)$$

where the last term corresponds to a summation over the continuum of the radiation modes of the flat metal/dielectric interface. In Section 3, we will provide an analytical expression for the continuum summation in the limit case of the scattering by infinitely small indentations and in the vicinity of the metal surface ($y<\lambda$). In Eq. (1), $\Psi_{SP}^+$ and $\Psi_{SP}^-$ respectively represent the electromagnetic fields of the forward– and backward–propagating SPPs, the analogue of the guided modes in waveguide theory. Hereafter, we assume that the SPP modes are normalized such that their magnetic fields are unitary on the interface at $x=y=0$. Formally, one may write for the forward–propagating SPP, $\Psi_{SP}^+ = [H_{z,SP}^+, E_{x,SP}^+, E_{y,SP}^+]$ with

$$\Psi_{SP}^+(x,y) = \exp(ikn_{SP}x) \begin{cases} \exp(ik\chi_d^{SP}y)\left[1, \dfrac{-\chi_d^{SP}}{\varepsilon_d}, \dfrac{n_{SP}}{\varepsilon_d}\right] \\ \exp(-ik\chi_m^{SP}y)\left[1, \dfrac{\chi_m^{SP}}{\varepsilon_m}, \dfrac{n_{SP}}{\varepsilon_m}\right] \end{cases}, \qquad (2)$$

where the top (resp. bottom) expression holds for $y>0$ (resp. $y<0$).

Similarly for the SPP propagating towards negative $x$, one obtain

$$\Psi_{SP}^-(x,y) = \exp(-ikn_{SP}x) \begin{cases} \exp(ik\chi_d^{SP}y)\left[1, \dfrac{-\chi_d^{SP}}{\varepsilon_d}, \dfrac{-n_{SP}}{\varepsilon_d}\right] \\ \exp(-ik\chi_m^{SP}y)\left[1, \dfrac{\chi_m^{SP}}{\varepsilon_m}, \dfrac{-n_{SP}}{\varepsilon_m}\right] \end{cases}, \qquad (3)$$

where $k=2\pi/\lambda$, $n_{SP} = [\varepsilon_m\varepsilon_d/(\varepsilon_m+\varepsilon_d)]^{1/2}$ is the normalized propagation constant ($n_{SP}=k_{SP}/k$) of the SPP mode, $\chi_m^{SP} = [\varepsilon_m-(n_{SP})^2]^{1/2}$ and $\chi_d^{SP} = [\varepsilon_d-(n_{SP})^2]^{1/2}$. In Eq. (1), $\alpha^+$ and $\alpha^-$ are the modal excitation coefficients of the two SPPs launched in opposite directions. They can be calculated by using the mode orthogonality [52]

$$\alpha^+ = (N_{SP})^{-1} \int_{-\infty}^{\infty} [E_y(x,z)H_{z,SP}^-(x,z) - H_z(x,z)E_{y,SP}^-(x,z)]dz, \qquad (4a)$$

$$\alpha^- = (N_{SP})^{-1} \int_{-\infty}^{\infty} [H_z(x,z)E_{y,SP}^+(x,z) - E_y(x,z)H_{z,SP}^+(x,z)]dz, \qquad (4b)$$

where the normalization constant $N_{SP}$ is given by

$$N_{SP} = \int_{-\infty}^{\infty} [E_{SP}^+(x,z)H_{SP}^-(x,z) - E_{SP}^-(x,z)H_{SP}^+(x,z)]dz = (i/k)(n_d^4-n_m^4)/(n_d^3 n_m^3). \qquad (5)$$

Note that the integral is piecewise constant, see Figs. 2c and 3. This can be easily derived by incorporating Eq. (1) into Eqs. (4a) and (4b) and by noting that the EH products systematically involve forward– and backward–propagating SPPs, such that the exponential $x$–dependence cancels out. Because we are concerned by metals with a finite conductivity, all these modes are not orthogonal in the sense of the Poynting–vector, as it is usually the case in waveguide theory with lossless materials, but they obey the unconjugate general form of orthogonality. That is the reason why EH products (instead of the usual EH* product) are used in Eqs. (4a) and (4b) [52–54]. For good metal ($|\varepsilon_m|>>\varepsilon_d$),

$$N_{SP} \approx -in_m/(k\, n_d^3), \qquad (6)$$

which is approximately four times the power flow of the normalized SPP modes of Eqs. (2) or (3). From Eqs. (4a) and (4b), the coefficients $\alpha^+$ and $\alpha^-$ are derived by numerically calculating the overlap integrals on the right–hand sides of the equations.

Figure 2c represents the modulus of the SPP scattering coefficients for the problem considered in Fig. 2b. The calculation has been performed for 500 values of $x$, $-2.5\lambda < x < 2.5\lambda$. In the slit section, the integrals defined in Eqs. (4a) and (4b) are meaningless (since the air−metal interface does not correspond to the plane $y=0$) and their calculated values are not presented for the sake of clarity. Let us consider $|\alpha^+|$, the discussion is similar for $|\alpha^-|$. On the right side of the aperture, the computed values for $|\alpha^+|$ are virtually all constant as evidenced by the blue line. Not only the modulus but also the argument of $\alpha^+$ is independent of $x$. This is actually consistent with the launching of a SPP mode $\{H_{SP}^+, E_{SP}^+\}$ on the right side of the groove. Consistently with the outgoing radiation condition at the groove, $|\alpha^+|$ is found to be nearly null (within numerical accuracy) on the left side of the aperture. In our opinion, all this provides a strong support for the soundness of the normal−mode−decomposition formalism used for quantifying the SPP excitation.

The above normal−mode theory decomposes the scattered fields as a superposition of normal modes. The SPP is the most important mode of the decomposition, since it is a bounded mode with a small characteristic damping. The picture that we emphasize in Fig. 2c corresponds to the launching of SPP modes on both sides of the slit. It is important to realize that the SPP scattering coefficients $\alpha^+$ and $\alpha^-$ can be defined unambiguously thanks to the normal−mode formalism. This vision that gives priority to the concept of normal modes largely contrasts with recent theoretical developments that mix the SPP mode with propagative and evanescent waves to propose the new concept of a transitory SPP with a complex−valued envelope [55−56]. We will not adopt this unusual approach hereafter, and will simply assume that a mode (such as the SPP), once launched, possesses a purely exponential dependence along the propagation direction.

The SPP−mode formalism is not restricted to the analysis of the SPP launching on semi−infinite planar interfaces in opposite directions. It can be as well applied to much more intricate structures, allowing for instance the calculation of the SPPs excited on a flat-interface section with a finite length. Figure 3 illustrates our purpose. The structure is composed of eleven grooves in a gold film. It is illuminated ($\lambda=800$ nm) by an incident Gaussian beam with a 5−$\lambda$ waist. The groove depths and widths have been optimized to maximize the energies of the SPPs launched on both sides of the groove array. For the optimal parameters, the SPP efficiency is as large as 69%, implying that two thirds of the incident photons are converted into two SPPs launched in opposite directions. The superimposed curves represent the modulus of the SPP scattering coefficients $\alpha^+(x)$ and $\alpha^-(x)$ calculated at every flat interface in between the grooves. Indeed because the illumination and the structure are symmetric with respect to the plane $x=0$, $\alpha^+(x) = \alpha^-(-x)$. Consistently with the modal picture, the coefficients $\alpha^+$ and $\alpha^-$ are constant in all the flat sections in between the grooves. It is remarkable to observe that, as one moves from the left to the right side of the structure, $|\alpha^+(x)|$ keeps on increasing and that the progressive increase is mainly due to the additional SPP launchings that occur at every individual groove. Note that this additional launching compensates for the SPP losses due to the SPP back−reflection and SPP out−of−plane scattering that also inevitably occurs at every individual groove [34].

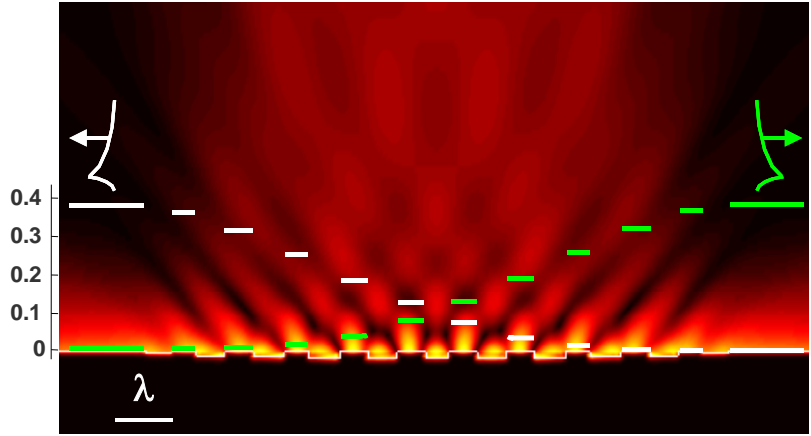

**Fig. 3.** SPPs scattered by a miniature optimized grating coupler in a gold substrate illuminated by a normally incident Gaussian beam. The image shows the scattered magnetic field, $|H_z|$. Note that the vertical and horizontal length scales are different. The incident Gaussian beam and the specularly reflected wave have been removed for the sake of clarity. The superimposed curves show the SPP generation strengths $|\alpha^+(x)|$ (green) and $|\alpha^-(x)|$ (white) obtained by calculating the overlap integrals of Eqs. (4) and (5). The calculation is performed for optimized groove depths and widths and for $\lambda=800$ nm ($n_m=0.18+5.13i$). The total coupler efficiency (ratio between the energy launched into the two SPPs and the incident energy of the Gaussian beam) is 69%.

## 2.2 SPP excitation efficiency

For practical applications, a fundamental issue concerns the launching (or decoupling) efficiencies of SPPs from freely propagating light. Conventional SPP couplers involve prisms or subwavelength–period gratings [6], and rely on interaction lengths that are comparable to the SPP damping characteristic lengths. In view of potential applications of such structures, it is essential to lower the interaction length and to reinforce the SPP interaction strength [57−61,25]. Thus a quantitative knowledge of the SPP excitation efficiency by individual nano−objects, like single discontinuities, slits, grooves or ridges, is a central problem for further engineering SPP−based devices [25−33]. In this section, we summarize the main known results.

Hereafter, the total SPP−excitation efficiency is defined as the ratio between the SPP power flux (launched on both sides of the indentation) and the incident power flux that directly impinges onto the indentation. This efficiency can be directly deduced from the modal excitation coefficients, $\alpha^+$ and $\alpha^-$, which are defined in Eq. (4a) and (4b). It is a matter of renormalization. The efficiency, denote by $\eta$ hereafter, may exceed unity for sub–$\lambda$ indentations, implying that more light than is geometrically incident upon the indentation is converted into SPPs. In general, $\eta$ depends on the incident wavelength, on the indentation shape and size, on the dielectric constant of the metal, and on the angle of incidence of the illumination beam. The latter is assumed to be a plane wave in the following.

Figure 4 summarizes the main useful results predicted for slits at an air−gold interface. Other noble metals used at frequencies close to the plasma frequency exhibit similar behavior. The results are obtained with the formalism developed in the previous Section over a broad spectral interval ranging from the visible to thermal infrared. Whether the slit aperture is illuminated by the fundamental slit mode (left side in Fig. 4) or by an incident plane-wave (right side in Fig. 4), the general trends are essentially the same. First, we note that the SPP excitation efficiency strongly depends on the slit width. It basically follows a universal curve that peaks at a value of the normalized slit width equal to $w_0 \approx 0.23\lambda$. Second, at visible wavelength ($\lambda=0.6$ μm, black curves), the efficiency is fairly large. When the incident light is incident on the interface from the slit side, the SPP excitation efficiency can become of order 0.5, implying that of the power coupled out of the slit half goes into heat, if no other corrugation is present to decouple the launched SPPs to radiation modes [62]. Third the efficiency rapidly decreases for large wavelengths; it is only 2.8% for λ=10 μm (blue curves).

The computational data can be quantitatively explained with an approximate model valid for small sub−λ slit widths [18,49]. The model provides a closed−form expression for the efficiency

$$\eta = (\varepsilon_d/|\varepsilon_m|)^{1/2}\, F(w/\lambda), \qquad (7)$$

where F($w/\lambda$) is proportional to a sinc−like function of the normalized slit width, which weakly depends on the dielectric constants of the problem. Equation (7) shows two important trends. It predicts that the plasmonic excitation can be enhanced by immersing the sample in a dielectric material (it increases proportionally to $n_d$), and that the efficiency scales as $|\varepsilon_m|^{-1/2}$ (as the inverse of the wavelength for a Drude model). This scaling law, see Section 4 for additional details, indicates that the visible part of the electromagnetic spectrum is the most exciting region for studying SPPs.

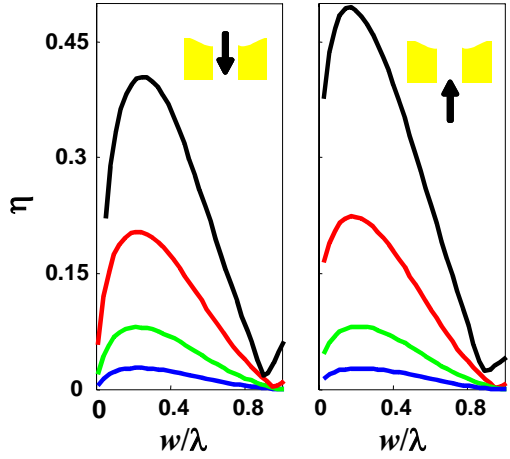

**Fig. 4.** Total SPP excitation efficiency by a sub−λ slit as a function of the normalized slit width $w/\lambda$ for different wavelengths. Left: the interface is illuminated by the fundamental mode of the slit. Right: The interface is illuminated by a normally incident plane wave. From top to bottom, the black, red, green and blue curves are obtained for λ=0.6, 1, 3 and 10 μm assuming the gold frequency-dependent permittivity $\varepsilon_m$ taken from [16]. The total efficiency corresponds to the efficiency of the two SPP modes launched in the two opposite directions on the air-gold interface. In agreement with Eq. (7), the SPP efficiency scales as $|\varepsilon_d/\varepsilon_m|^{1/2}$. More details can be found in [18] and [49].

The sinc−like behavior of the SPP−excitation efficiency has been verified experimentally. A qualitative agreement has been obtained for the efficiency of SPPs excited by the slit mode [63]. In the experiment performed at λ=780 nm, the SPPs launched at the transmission facet of a single slit are detected with a near−field probe positioned far away from the slit to avoid the contamination of quasi-CWs. Essentially, the recorded SPP−efficiency follows a sinusoidal variation with the slit width, the maximum efficiency being ten times larger than the minimum. By analyzing the far-field optical images recorded in the Fourier plane of a leakage microscopy setup, SPP−launching efficiencies in reasonably good agreement with the predictions of Fig. 4 have been measured [64]. More recently, a quantitative agreement showing a maximum and almost-null efficiencies for $w_0 \approx 0.23\lambda$ and $w_0 \approx \lambda$, respectively, has been obtained with far−field measurements performed on a slit doublet when only a single slit is illuminated at λ=810 nm [20].

There are two ways to increase the SPP−generation efficiency of slits. In qualitative agreement with computational results, a 64%−efficiency has been obtained with near−field measurements [48] performed at λ=975 nm for a gold nanoslit illuminated at a highly−oblique incidence of 75°. The value of 64% represents a 3−fold enhancement in comparison with the maximum efficiency of 20% predicted for λ=1 μm (red curve) in Fig. 4 at normal incidence. The SPP−generation efficiency can be also increased by considering grooves instead of slits. As shown by computational results [50], a large fraction of the incident energy can be converted into SPPs under highly oblique incidence. Additionally, an almost−unidirectional SPP−launching may be achieved for certain incidences and for slit width larger than λ/2, a value above which the first anti−symmetric slit mode becomes propagative. Figure 5 shows the SPP−excitation efficiencies as a function of the normalized groove depth $h/\lambda$ for λ=0.8 μm (solid−black curve) and λ=1.5 μm (dashed−dot red curve). For Fabry−Perot groove resonances, the light converted into SPPs is two to three times larger than that which is directly incident onto the aperture area.

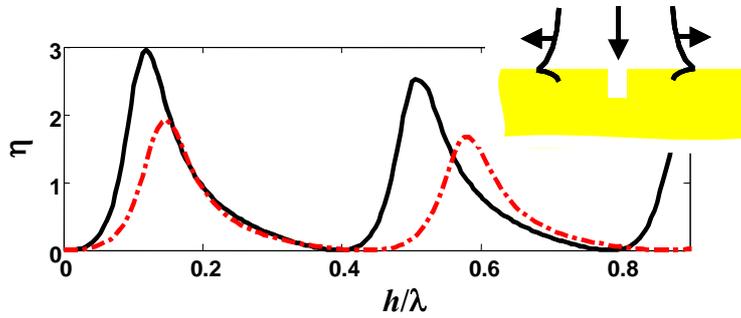

**Fig. 5.** Total SPP−excitation efficiency (in the two opposite directions, see the inset) for a sub−λ groove in a gold substrate as a function of the normalized groove depth $h/\lambda$. The data are calculated with the a-FMM for a groove width $w$=0.1λ. The results hold for gold and for normal incidence from air ($\varepsilon_d$=1). The solid-black and dashed−dot−red curves are obtained for λ=0.8 μm and 1.5 μm, respectively. Maximum excitation efficiencies are obtained for Fabry-Perot resonances. More details, including an analytical treatment, can be found in [49].

Apart from slits and grooves, dielectric or metallic ridges on metal films also deserve attention. Figure 6 shows the SPP−excitation efficiency measured for gold ridges with a small 50−nm height [65]. The ridges are illuminated under normal incidence by a focused laser beam (λ=800 nm) and the launched SPPs are detected by their leakage radiations in the far−field region [66−67]. For these tiny ridges, the SPP−generation efficiency can be remarkably improved by arraying the ridges like in a grating−SPP coupler. The efficiency increase is however achieved at the expense of miniaturization. Theoretical analyses [50] predict that, for a comparable object volume, ridges (even dielectric ones) may provide SPP−excitation efficiencies that exceed those obtained for slits and grooves at a given wavelength. Another way to improve the efficiency consists in using metallic ridges on dielectric posts [50]. This gap geometry is theoretically predicted to provide higher SPP excitation efficiencies since it combines several effects that favor the exaltation of resonances.

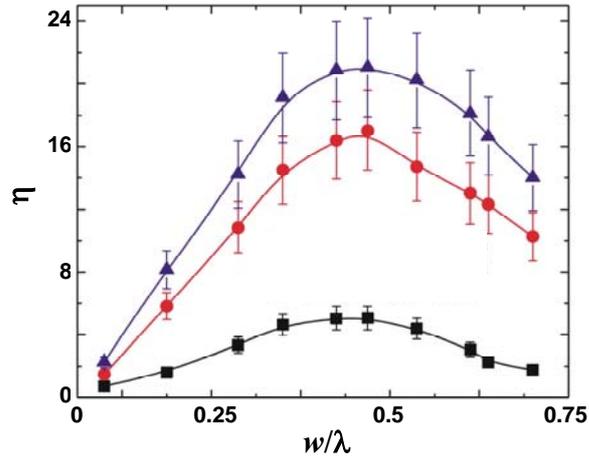

**Fig. 6.** Total SPP−excitation efficiency of tiny 50−nm−high ridges with different widths. Three curves show three sets of measurements performed on one (black squares), three (red circles) and five (blue triangles) ridges aligned with a periodicity equal to the wavelength (λ=800 nm). More details can be found in [65].

## *3. Generation of quasi−cylindrical waves*

The field scattered by nanoparticles in the vicinity of metallic surfaces is not solely composed of SPPs. This statement has been convincingly addressed by initial studies on the radiation from a Hertzian dipole in the presence of a boundary between two quite different materials, like air and conductors. This scattering problem has been of long-standing interest in classical electromagnetism. In particular for long−distance radio transmission, the Hertzian dipole radiation has been a controversial subject over many years in relation with the existence or nonexistence of the Zenneck surface mode [68−69], the analogue of the SPP in cylindrical coordinates with a $r^{-1/2}$ rate of decrease. The interested reader can refer to the recent review article by Collin [70], which well explains the subtle mathematics required for solving the radiation problem and well summarizes some errors that have been recurrently made in the analysis over many years, since the initial 1909's treatment by Sommerfeld.

Curiously, the idea that the near−field launched on a metallic surface by sub−λ indentations may significantly differ from a pure SPP mode has been only recently considered at optical frequencies. So far, only 2D indentations have been studied in details and in a confusing manner. The existence and importance of a wave (different from the SPP) have been initially promoted by experimental data involving the interference between the surface waves launched from a sub−λ groove and the transmitted light through a neighboring slit [71]. The experiment revealed the existence of a short−range transient regime, in which the surface waves have a rapid drop in amplitude within the first few micrometers from the groove, before reaching a long−range regime. The interpretation was given using a composite diffracted evanescent waves model (CDEWs) [71−72], which predicts an amplitude drop as $1/x$ with the distance *x* from the groove. It has been rapidly challenged by numerical results and by a theoretical analysis suggesting that the launched surface field consists of a long−range SPP−mode and of a short−range "creeping wave", which was computed numerically [37] and which is *called a quasi−cylindrical wave in the present work*. This first theoretical analysis has been confirmed by experimental near−field measurements obtained for a slit Young's doublet experiment [36], showing that the near−field distribution between the slits is composed of two sets of counter−propagative waves, each formed by the superposition of a SPP and a rapidly−decaying wave with a free−space propagation constant. These initial works have been pursued by theoretical contributions aiming at better understanding the properties of the quasi-CW. In an attempt to find analytical solutions,

Lévêque and his collaborators [73] used a simplified model, artificially inserting two poles in the scalar theory based on the CDEW in order to retrieve the SPP contribution. This empirical operation based on physical intuition is however not rigorously valid. In more recent works, Sheng and his collaborators [55–56] use the method in [70] and introduce a transient SPP, which asymptotically damps as $-\ln(x)$, $x^{-1/2}$ and $x^{-1}$, respectively in different distance ranges from the magnetic line source. In our opinion, this analytical treatment is not sound physically; hereafter fully–computational and analytical results both evidence two regimes with attenuation varying as $x^{-1/2}$ and $x^{-3/2}$.

The analysis performed in this Section is an attempt to clear up the confusing situation. We start by providing a numerical example for the quasi-CW by considering the scattering by a sub–λ slit. Then in Section 3.2, we consider the Green-function response of a dipole line source located just above or just below a metallo–dielectric interface. The analysis is similar to that in [37], but in addition we provide an analytical expression for the branch–cut integral that was performed numerically in the former work. The analysis represents an *original contribution* of the present report. Most of the mathematical details are moved to the Appendix A for the sake of clarity. In our opinion, the analytical work strongly clarifies the properties of the quasi-CW that are discussed in the sub-Section 3.3.

## 3.1 Field scattered by a sub–λ slit

To illustrate our purpose, let us consider the scattering problem shown in Fig. 7a, where a gold–air interface is illuminated by the fundamental mode of a slit. For the sake of generality, a small indentation is considered on the right side of the slit, removing the vertical symmetry of the problem. Figure 7b shows the total magnetic field scattered in air, and Fig. 7c shows the SPP contribution to the total field, obtained by applying the formalism developed in the previous section. In Fig. 7d, we show the difference between the total and the SPP fields. Above the interface in the far field, this residual wave (like the total field) looks like a cylindrical wave. This is a direct consequence of the 2D geometry. Spherical-like waves would be obtained for local (hole for instance) defects on the surface.

The knowledge of the optical characteristics of the residual wave in the vicinity of the metallic surface is crucial for understanding the optical properties of corrugated metallo-dielectric surfaces. It is this field, together with the SPP–mode field, which are responsible for the electromagnetic interaction between the sub–λ indentations on the surface. At first glance, the residual wave may appear to possess a complex behavior, especially in the vicinity of the surface. Additionally for large indentations with transversal dimensions greater than the illumination wavelength, it may significantly vary from one geometry to the other. However, as long as we are concerned by sub–λ indentations, as in the present work, it is possible to obtain an intuitive and helpful description of the residual wave (see the following sections 3.2 and 3.3).

Hereafter, *the residual wave scattered by 2D infinitely–small indentations will be called a quasi–cylindrical wave.* By "sub–λ", we do not refer to very weak scatterers well predicted by perturbative theories. Indeed, as will be illustrated hereafter (see Fig. 10 in particular), the residual scattered field of micro/nano–indentations that incorporate localized resonances and that are currently of great importance for plasmonic and metamaterial devices can be accurately modelled as a quasi-CW obtained from the asymptotic limit of 2D *infinitely–small indentations*.

Figure 7e shows the magnetic field scattered on the surface ($y$=0). The total field (solid curve) is composed of a SPP (dotted curve) contribution and of a quasi-CW (dashed curve) contribution. In general, the quasi-CW attenuation rate is much faster than the characteristic damping of the SPP mode. As shown in Section 4, this is true for noble metal over the full electromagnetic spectrum, from visible to THz frequencies. What changes with the excitation energy is the initial excitation rates of the SPP and of the quasi-CW.

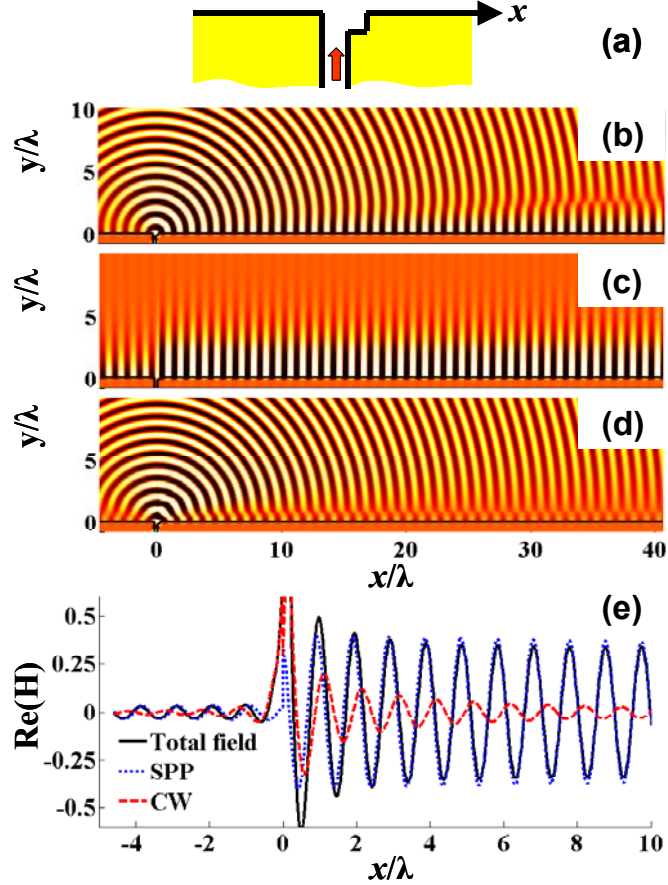

**Fig. 7.** Electromagnetic field scattered by a sub–λ slit in a gold substrate under illumination by the fundamental slit mode. (a) Sketch of the scattering problem. The slit width is $w=.25\lambda$ ($\lambda=800$ nm). The small indentation (with a 0.187λ–width and a 0.187λ–depth) on the right side of the slit is used to remove the vertical symmetry and to show a general example, in which the scattered fields are different for $x>0$ and $x<0$. (b) Total magnetic field scattered in air. (c) Magnetic field associated to the scattered SPP modes. The SPP contributions have been extracted from the total field by using the formalism developed in Section 2. Note that, due to the small indentation, the SPP generation strength on the right side of the slit is much larger than that on the left side. (d) Difference between the total magnetic field and the SPP field. The quasi-CW is defined as the residual contribution on the surface. (e) Magnetic field on the surface ($y=0$). Total field (solid–black), SPP-mode contribution (dotted–blue) and quasi-CW contribution (dashed–red). The calculations are performed for an incident field with a unitary magnetic field at $x=y=0$.

## 3.2 Analytical solution of the line-source problem

In the orthogonal Cartesian coordinate ($x,y,z$) system, we consider an interface $y=0$ between two semi-infinite media with relative permittivities $\varepsilon_d$ (dielectric) and $\varepsilon_m$ (metal). Under transverse magnetic polarization, the Green−function response of a dipole line source located at the point $x=y=0$ satisfies the following equations

$$\nabla\times\mathbf{E} = ik\mathbf{H} + H_z^S\mathbf{z}\,\delta(x,y) \qquad (8)$$

$$\nabla\times\mathbf{H} = -ik\varepsilon(\mathbf{r})\mathbf{E} + (E_x^S\mathbf{x} + E_y^S\mathbf{y})\,\delta(x,y), \qquad (9)$$

where **x**, **y** and **z** are the unitary vectors along the $x,y$ and $z$ axes, $\delta(x,y)$ is the 2-D Dirac distribution, and **E**=$E_x$**x**+$E_y$**y** and **H**=$H_z$**z**. The superscript S is used to denote the source currents. Note that, although electric sources only have to be considered since nonmagnetic materials are considered here, we have introduced a magnetic line source in Eq. 8 for the sake of completness. The relative permittivity $\varepsilon(\mathbf{r})$ is a discontinuous function of the space coordinate: $\varepsilon(x,y)=\varepsilon_m$ for $y<0$ and $\varepsilon(x,y)=\varepsilon_d$ for $y>0$.

In the Appendix A, we establish an asymptotic closed–form solution of Eqs. (8) and (9) using a modified steepest descent method that takes into account the presence of a neighboring pole. Our analysis is related to that in [70]. The SPP contribution to the radiated field is seen as a pole in the formalism and is fully analytical. The treatment of the quasi-CW is much more intricate. We have developed an approximate expression that is quantitatively accurate for $|x|>\lambda$ in the vicinity of the metal surface ($|y|<\lambda$). With these approximations, the field radiated by the line source, $\Psi(x,y)=[H_z, E_x, E_y]$, can be decomposed as a sum of four waves, two SPPs and two quasi-CWs, which are launched in opposite directions. Formally, one may write

$$\Psi(x,y) = [\alpha^+ \Psi_{SP}^+ (x,y)+\alpha^- \Psi_{SP}^- (x,y)]+ [\beta^+ \Psi_{CW}^+ (x,y)+\beta^- \Psi_{CW}^- (x,y)], \qquad (10)$$

where $\Psi_{SP}^+=[H_{z,SP}^+, E_{x,SP}^+, E_{y,SP}^+]$ and $\Psi_{SP}^-=[H_{z,SP}^-, E_{x,SP}^-, E_{y,SP}^-]$ are the normalized ($H_{z,SP}(0,0)=1$) SPP modes that are given by Eqs. (2) and (3) and that are launched in the positive and negative $x$–directions. $\alpha^+$ and $\alpha^-$ represent the SPP–excitation coefficients. In the analytical treatment, they are obtained as pole–residues

$$\alpha^\pm = -[H^S + \chi_m^{SP}/\varepsilon_m E_x^S \pm n_{SP}/\varepsilon^S E_y^S]/N_{SP}. \qquad (11)$$

where $\varepsilon^S$ is either $\varepsilon_d$ or $\varepsilon_m$, whether the Dirac source is located in the dielectric material or in the metallic medium.

In Eq. (10), $\Psi_{CW}^+=[H_{z,CW}^+, E_{x,CW}^+, E_{y,CW}^+]$ and $\Psi_{CW}^-=[H_{z,CW}^-, E_{x,CW}^-, E_{y,CW}^-]$ are the quasi-CWs, which are respectively launched in the positive and negative $x$–directions. A very accurate analytical expression for these waves is derived in the Appendix A. We have

$$\Psi_{CW}^+ (x,y) \approx \frac{W(2\pi\gamma|x|/\lambda)}{W(2\pi\gamma)} (\lambda/|x|)^{3/2} \Psi_0^+ (x,y), \qquad (12a)$$

for $x>0$, and 0 otherwise. Similarly for the quasi-CW launched in the opposite direction,

$$\Psi_{CW}^- (x,y) \approx \frac{W(2\pi\gamma|x|/\lambda)}{W(2\pi\gamma)} (\lambda/|x|)^{3/2} \Psi_0^- (x,y), \qquad (12b)$$

for $x<0$, and 0 otherwise. In Eqs. (12a) and (12b),

$$\gamma = n_{SP} - n_d = -n_d^3/(\varepsilon_d + \varepsilon_m + n_m^3/\chi_m^{SP}) \qquad (13)$$

and W(t) is an Erf–like envelop function of the complex variable t, see Fig. 18 in the Appendix A. W(t) governs the decaying rate of the quasi-CW. For the following, it is sufficient to know that W(t) slowly varies with t, W(t) $\propto$ 2it for $|t|<<1$ and W(t) tends towards unity as $|t|\to\infty$. The vectors $\Psi_0^+$ and $\Psi_0^-$ in Eqs. (12a) and (12b) provide the vector–field pattern of the quasi-CWs. As expected intuitively, they correspond to the electromagnetic–field solutions of Maxwell's equations when a metallic surface is illuminated from the dielectric medium by a plane wave at grazing incidence ($k_{//}=\pm k n_d$) under TM polarization. For a normalized ($H_z(0,0)=1$) field and for $y>0$, $\Psi_0^+$ and $\Psi_0^-$ are given by

$$\Psi_0^\pm(x,y) = \exp(\pm ikn_d x)\,[1-iky(\varepsilon_d \chi_m^0/\varepsilon_m),\ \chi_m^0/\varepsilon_m,\ \pm(1-iky(\varepsilon_d \chi_m^0/\varepsilon_m))/n_d], \tag{14a}$$

and for $y<0$ by

$$\Psi_0^\pm(x,y) = \exp(\pm ikn_d x)\,\exp(-ik\chi_m^0 y)\,[1,\ \chi_m^0/\varepsilon_m,\ \pm in_d/\varepsilon_m], \tag{14b}$$

with $\chi_m^0 = (\varepsilon_m - \varepsilon_d)^{1/2}$. Note the remarkable linear behavior with the $y$ variable of the $H_z$ and $E_y$ field–components in Eq. (14a). Similar expressions hold for $\Psi_0^-$ with an $\exp(-ikn_d x)$ dependence.

Thus the quasi-CW admits *a simple interpretation*: it is the product of the asymptotic and intuitive grazing field by an envelope function.

Finally, let us provide the closed–form expressions for the CW–excitation coefficients, $\beta^+$ and $\beta^-$. They are given by (see Appendix A)

$$\beta^\pm = k\,\frac{\varepsilon_m^2 \sqrt{i n_d}}{\varepsilon_d(\varepsilon_d - \varepsilon_m)}\,\frac{W(2\pi\gamma)}{4\pi^2}\,[H^s + \chi_m^0/\varepsilon_m E_x^s \mp n_d/\varepsilon^s E_y^s]. \tag{15}$$

It is worth mentioning that, since $\Psi_{CW}^+$ and $\Psi_{CW}^-$ diverge for $x \to 0$, we have decided to normalize the quasi-CW magnetic field such that $H_{z,CW}^+(\lambda,0) = H_{z,CW}^-(-\lambda,0) = 1$. Moreover, because the SPP modes are normalized such that their magnetic field is unitary at $x=y=0$, the ratio $|\beta^+/\alpha^+|$ (or $|\beta^-/\alpha^-|$) conveniently quantify the relative strength between the SPP and quasi-CW generated at a one–wavelength distance from the indentation.

## 3.3 Properties of the quasi-CW

The purpose of this subsection is to present the important characteristics of the quasi-CW, such as the decaying rate, the phase velocity … Additionally, we check the accuracy of approximate model by comparison with fully–vectorial computational data obtained either for the exact solution of the line source problem or for a finite-dimension sub–$\lambda$ scatterer on a gold surface, namely the slit with a small attached indentation in Fig. 7.

Before going into the quasi-CW properties, it is important to realize the approximations made with the asymptotic closed–form solution of Eqs. (12a) and (12b). Figure 8 compares the electromagnetic fields generated by a dipolar line source, calculated without approximation (left–hand panel) or by considering the closed–form expressions (right–hand panel). The figure holds for a line source polarized along the $x$–direction ($E_x^S=1$) and located just above the metal surface at $x=y=0$, for $n_d=1$ and $n_m=0.18+5.13i$ (gold at $\lambda=800$ nm). Figures 8a and 8b show the total magnetic field. We immediately understand that the cylindrical character of the total field above the surface is not taken into account by the asymptotic expansion used in the model for grazing incidence. The latter is only accurate in the near–field just above the interface. Figures 8c and 8d are identical and correspond to the SPP magnetic–field contribution. The six lowest figures 8e–8j correspond to the electromagnetic–field components of the quasi-CW. Again let us note that the asymptotic model is accurate only in the near−field of the interface, including the unexpected fact that the main field components, $H_{z,CW}$ and $E_{y,CW}$, are almost null in a horizontal plane located just above the interface.

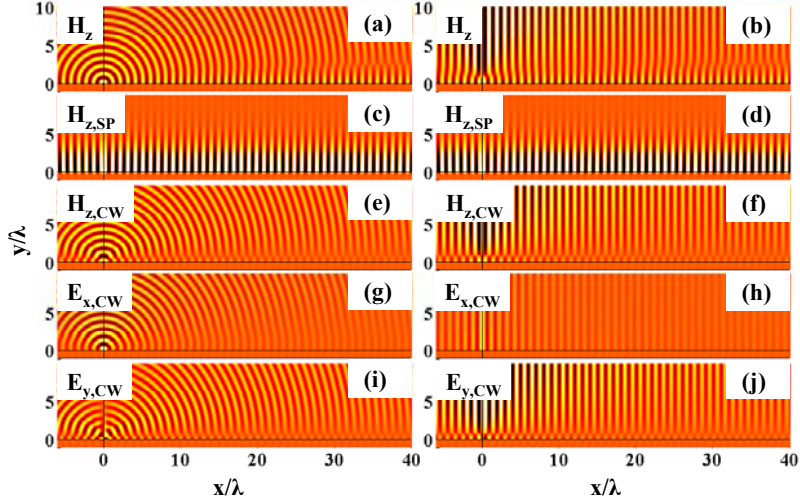

**Fig. 8.** Electromagnetic fields generated by a line source polarized along the *x*-direction and located just above the metal surface at $x=y=0$. The calculation is performed for $n_d=1$ and $n_m=0.18+5.13i$, which corresponds to gold at $\lambda=800$ nm. The left part shows the exact solution obtained by numerically solving Eqs. (A3) and (A4), see the appendix. The right side of the figure shows the solution obtained with the approximate model. (a) and (b) Total magnetic field. (c) and (d) SPP magnetic-field contribution. (e) and (f) quasi-CW magnetic−field contribution. (g) and (h) quasi-CW electric-field (*x* polarization) contribution. (i) and (j) quasi-CW electric-field (*y* polarization) contribution.

### 3.3.1 Asymptotic decaying rates

A quantitative knowledge of the quasi-CW decaying rate in the *x*−direction is crucial since it is this field that is responsible for the electromagnetic interaction between sub−λ particles on corrugated metal films. With the approximate model, the decaying rate is provided by the envelope function in Eqs. (12a) and (12b). Two different regimes are predicted. For $\gamma kx \ll 1$, since $W(t) \approx 2it$ for small $|t|$, the envelope function $W(2\pi\gamma|x|/\lambda) (\lambda/|x|)^{3/2}$ scales as $(\lambda/|x|)^{1/2}$. The inverse square root dependence is rather expected for this 2D problem and has already been discussed in all related works [37,55−56]. Additionally, let us note that for large metal conductivities, since the parameter $\gamma$ in Eq. (13) vanishes as $-n_h^3/\varepsilon_m$, the approximate expression $W(t) \approx 2it$ becomes valid for $|x|>\lambda$. This is consistent with our expectation that, for the line−source emission over a perfect conductor sheet, the emitted field is a cylindrical wave with an asymptotic attenuation behavior $|x|^{-1/2}$, even for very large *x*'s. More unexpected is the quasi-CW scaling for large *x*'s and for a finite conductivity. In this case, $W(2\pi\gamma|x|/\lambda)=1$ and the envelope function decreases as $|x|^{-3/2}$. To summarize, one gets

$$|\Psi_{CW}^+(x,0)| \propto x^{-1/2} \text{ for } kx \ll |\varepsilon_m|, \tag{16a}$$

$$|\Psi_{CW}^+(x,0)| \propto x^{-3/2} \text{ for } kx \gg |\varepsilon_m|. \tag{16b}$$

The attenuation behavior with the cube of the inverse square root for large *x*'s is a surprising prediction of the analytical model. It is exact since the analytical model becomes rigorous for $x/\lambda \to \infty$. It has not been discussed in earlier works [37] that reported a faster decaying–rate at large $x$'s and is not consistent with the predictions in [55–56]. In Fig. 9, we quantitatively check the decaying–rate predictions of Eqs. (16a) and (16b). The dashed–green curves show fully–vectorial results obtained with the a–FMM for the scattering problem of Fig. 7a (an interface illuminated by the fundamental mode of an indented slit). The calculations have been performed for $\lambda=0.8$ μm (a) and $\lambda=3$ μm (b), respectively. The red–solid curve is obtained with the closed–form expression of Eq. (12a). Note that the amplitudes $E_x^S$ and $E_y^S$ ($H^S=0$) of the line source have been fitted so that the scattering coefficients $\alpha^+$ and $\alpha^-$ of Eq. (10) equal those calculated for the indented slit. The very good agreement (except for $|x|<<\lambda$) between the model predictions and the fully–vectorial calculation results well supports the scaling behaviors shown with the two dashed–dot black lines.

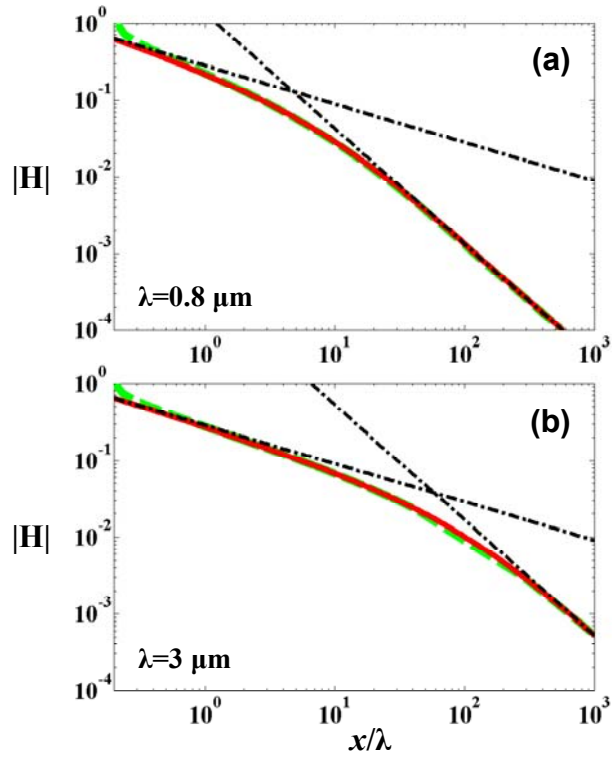

**Fig. 9.** Decaying rate of the quasi-CW as a function of the distance from the scatterer. The green-dashed curves show the modulus $|H(x,0)|$ of the magnetic field scattered on the right side ($x>0$) of the indented slit for the scattering problem of Fig. 7a. This fully–vectorial data are obtained with the a-FMM. The solid-red curve is obtained with the approximate expression for an electric line source. (a) $\lambda=800$ nm, $n_m=0.18+5.13i$. (b) $\lambda=3$ μm, $n_m=1.64+18.59i$. The asymptotic scaling behaviors for $x << |\varepsilon_m|\lambda$ and for $x >> |\varepsilon_m|\lambda$, Eqs. (16a) and (16b), are shown with the two dashed-dot black lines.

Figure 10 shows the quasi-CW amplitude scattered on the metal surface either in the positive or in the negative $x$–directions. Again an excellent agreement over the entire spatial interval ($|x|<10\lambda$) is found between the a–FMM numerical data (solid–black curve) and the closed–form expression of Eq. (13a) (dashed–dotted–green curve). The dashed–red curve shows a fit using the convenient expressions

$H_{z,CW}(x>0,y=0) = H_+ (x/\lambda)^{-m} \exp(ikn_d x)$, (17a)

$H_{z,CW}(x<0,y=0) = H_- (-x/\lambda)^{-m} \exp(-ikn_d x)$, (17b)

where H$_+$ and H$_-$ are two constants and *m* is a damping parameter equal to 0.872 and obtained by fit. The simple *m*–damping expression (*m* depends on the metal permittivity) has been used in previous works [36–37]. Although less accurate than the approximate closed–form expressions (see the right–hand inset), it well approximates the intermediate regime and is rather convenient and quantitative for intermediate distances ($|kx|<|\varepsilon_m|$).

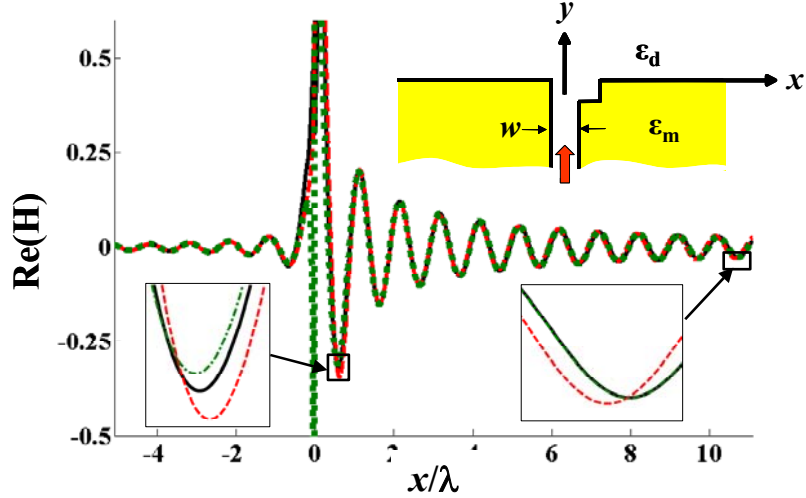

**Fig. 10** Quasi-CW magnetic–field scattered on the metal surface (*y*=0) in the vicinity of the scatterer ($|x|<10\lambda$). The results are obtained for the indented slit. Solid–black curve: Fully–vectorial data obtained with the a-FMM. Dashed–dotted–green curve: closed-form expression obtained from Eq. (13). Dashed–red curve: fit using the simple $x^{-m}$ expression of Eq. (17) obtained for *m*=0.872. The calculation is performed at $\lambda$=800 nm.

### 3.3.2 Normal dependence of the quasi-CW field

Another interesting property of the quasi-CW is that the magnetic and electric–*y*–components of the quasi-CW fields, $H_{z,CW}$ and $E_{y,CW}$, are almost null just above the metallic surface, see Figs. 7d and 8. Note that this property does not hold for the *x*–component $E_{x,CW}$ of the electric field. From Eqs. (14a) and (14b), one easily finds that the fields are nearly null for

$$y_0 = \lambda \, \Re\mathrm{e}\left(-i\varepsilon_m/(2\pi\varepsilon_d \chi_m^0)\right). \tag{18}$$

For noble metals at visible frequencies, $y_0$ corresponds to a sub–$\lambda$ distance even for $\varepsilon_d$=1. Since $H_{z,CW}$ and $E_{y,CW}$ are the dominant fields, $|E_{y,CW}|>>|E_{x,CW}|$, the quasi-CW appears as an electromagnetic field that seems to "stick" on the metallic surface in a wavelength scale.

### 3.4 The hybrid–wave concept

It is very important to note that for good metals, since $|n_m|>>|n_d|$, $\chi_{SP} \approx \chi_m$ and $n_{SP} \approx n_d$, the last square–bracket terms in Eqs. (11) and (15) become identical. Therefore, the $\alpha$ and $\beta$ coefficients become proportional for good metals *for an arbitrary line–source polarization*, i.e. for any $H^s, E_x^s, E_y^s$ triplets. Let us denote by $\rho$ the ratio, $\rho=\beta^+/\alpha^+=\beta^-/\alpha^-$, $\rho$ being approximately equal to $(-i)^{1/2}(n_m^3/n_d^{5/2})W(-\pi n_d^3/n_m^2)/(4\pi^2) \approx (-i)^{1/2}n_m/(2\pi n_d^{3/2})$ for good metals.

This implies that the field radiated by the line source, $\Psi(x,y)=[H_z, E_x, E_y]$, which is in general a sum of four independent waves, see Eq. (10), can be reduced to a sum of two waves that are launched in opposite directions. These waves will be called *hybrid waves* hereafter. Formally, one may write

$$\Psi(x,y) = \gamma^+ \Psi_{HW}^+ (x,y) + \gamma^- \Psi_{HW}^- (x,y), \tag{19}$$

where $\Psi_{HW}^+ = (1+\rho)^{-1}[\Psi_{SP}^+ + \rho \Psi_{CW}^+]$ and $\Psi_{HW}^- = (1+\rho)^{-1}[\Psi_{SP}^- + \rho \Psi_{CW}^-]$ are normalized such that the magnetic field is almost unitary at a single–wavelength distance from the source. Equation (19) is very important. In particular, when a sub–λ indentation is illuminated by an incident field, it can be assumed to respond with an induced dipole determined by its electric polarization tensor. The induced dipole **p**=[$p_x$, $p_y$] is difficult to evaluate in general (an example is provided in Section 4), but since the $x$ and $y$ dipoles generate the same hybrid wave on the surface, the overall shape of the scattered field is always the same for a given wavelength and a given metal. Only the amplitudes of the scattered fields, the $\gamma^+$ and $\gamma^-$ in Eq. (19), change with the incident illumination and with the specific shape of the indentation. This property will be used in Section 5 to derive a generalized formalism for analyzing sub–λ metallic surfaces.

It is interesting to consider the dominant contribution to the hybrid wave. For large $x$'s, because of the exponential damping of the SPP, the quasi-CW dominates. At intermediate distances from the source, the SPP largely dominates, but for small $x$'s, because it diverges, the quasi-CW field is much larger than the SPP field. The equality occurs for

$$x_0 = \lambda \left| \frac{(n_m^2 + n_d^2)^2}{2\pi n_m^2 n_d^3} \right| \approx \lambda \left| \frac{n_m^2}{2\pi n_d^3} \right|, \tag{20}$$

an expression that is easily derived from Eqs. (11) and (15), and that is consistent with FDTD computational results obtained in [74].

## *4. Scaling law*

So far, we have considered the SPP and quasi-CW generation rates for a single wavelength. We now investigate how the rates scale with the energy of the incident photon, emphasizing the key role played by the metal dispersion.

The problem at hands is illustrated in Fig. 11. The left side shows a sub–λ indentation located in the vicinity of a metal surface and illuminated by an incident plane wave (wavelength λ) with a unitary magnetic field at $x=y=0$, $H_{inc}(0,0)=1$. The right part of the figure illustrates a related scattering problem, where all the geometrical dimensions and the incident wavelength (λ'=Gλ) are all scaled by the same factor G. We assume that the incident field is kept unchanged by the scaling transform, so that $H'_{inc}(0,0)=1$. The questions we have in mind are for instance: what is the relationship between the scattered field $\psi_s = \{\mathbf{E}_s(x,y), \mathbf{H}_s(x,y)\}$ at λ and that at λ'? How do the relative fractions of the launched SPPs and quasi-CWs scale as the energy of the incident photon decreases?

The questions admit a very simple answer provided that we assume that the metal is not dispersive ($\varepsilon_m' = \varepsilon_m$). With this hypothesis, Maxwell's equations directly guaranty that, for this two–dimensional problem, the scattered field $\psi_s'$ at λ' is identical to that at λ, i.e. $\psi_s'(Gx,Gy) = \psi_s(x,y)$. However metals are highly dispersive materials and the question arises how the dispersion influences the problem.

There is no general solution to the questions for arbitrary λ and λ'. For instance, if we consider gold as the noble metal, it is not possible to infer what is the SPP–field scattered at λ'=3 μm (for instance) from the knowledge of the scattered field at any other smaller wavelength. However is it possible to predict how this scattered field scales as λ and λ' both tend to infinity (long wavelength regime). From what wavelength one may consider that the long-wavelength regime is reached is difficult to say. This probably depends on the incident illumination and on the size and shape of the sub−λ indentations. However, as we shall see, in practice the scaling law is qualitatively verified for important geometries even for small wavelengths in the near−infrared.

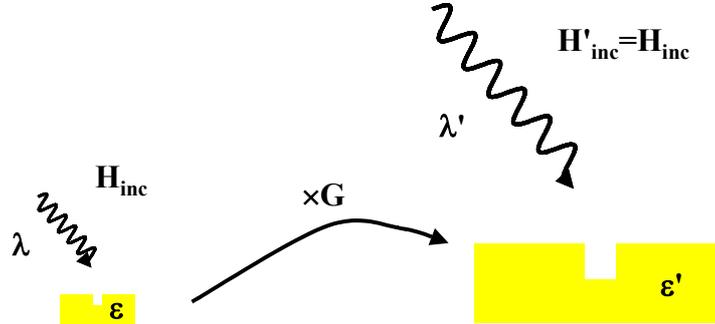

**Fig. 11** Scaling of the SPP and quasi-CW excitation coefficients as the incident wavelength is multiplied by a factor G and as all the geometrical parameters are scaled accordingly. Note that the incident field is kept unchanged. In general if $\varepsilon_m$' and $\varepsilon_m$ are not related, there is no way to infer any scaling law between the two scattering problems. However for noble metals, because the metal permittivity diverges as the wavelength increases ($|\varepsilon_m|\propto\lambda^2$ according to the Drude model), it is possible to infer an asymptotic behavior for the excitation coefficients, that is accurately satisfied even near−infrared wavelengths and beyond.

## 4.1 Perturbative approach

A natural approach to derive the scaling behavior consists in assuming that the sub−λ indentation is very small compared to the wavelength and in using the analytical solution of the line−source problem in Section 3.3. Let us explore this approach in more details. Referring to the left side of Fig. 11, we denote by [**E$_2$ H$_2$**] the field that satisfies Maxwell's equations, $\nabla\times\mathbf{E_2} = ik\mathbf{H_2}$, $\nabla\times\mathbf{H_2} = -ik\varepsilon(r)\mathbf{E_2} - ik\Delta\varepsilon(r)\mathbf{E_2}$, where $\Delta\varepsilon(r)$ is the permittivity contrast induced by the presence of the indentation. Similarly, let us consider another electromagnetic field distribution [**E$_1$ H$_1$**] that satisfies Maxwell's equations in the absence of indentation (flat interface) for the same incident illumination, $\nabla\times\mathbf{E_1} = ik\mathbf{H_1}$ and $\nabla\times\mathbf{H_1} = -ik\varepsilon(r)\mathbf{E_1}$. Thus the scattered field, $\mathbf{E}=\mathbf{E}_2-\mathbf{E}_1$ and $\mathbf{H}=\mathbf{H}_2-\mathbf{H}_1$, obeys the classical equations

$$\nabla\times\mathbf{E} = ik\mathbf{H}, \quad \nabla\times\mathbf{H} = -ik\varepsilon(r)\mathbf{E} - ik\Delta\varepsilon(r)\mathbf{E_2}, \tag{21}$$

showing that the indentation acts as an electric surface source $-ik\Delta\varepsilon(r)\mathbf{E_2}$.

Under the assumption that the indentation dimensions are much smaller than the wavelength, we may further consider that the indentation acts as a line (Dirac) electric source $[H^s, E_x^S, E_y^S] = -ik\Delta\varepsilon(r)S[0, E_{x,2}, E_{y,2}]$, where S is the *effective* indentation area and $E_{x,2}$ or $E_{y,2}$ represents the actual field distribution *that is assumed to be constant* on the effective area. According to Eq. (11), it is easily seen that for inclusions in the dielectric material (ridges for instance) the $E_y^S$–contribution that scales as $1/\varepsilon_d$ dominates, whereas for inclusion in the metal (grooves for instance), it is the $E_x^S$–contribution ($\propto|n_m|^{-1}$) that dominates.

The difficulty arises when considering the values of the actual field distributions in the inclusion, $E_{x,2}$ and $E_{x,2}$. Naively, one may apply a first−order Born approximation and replace the actual field $\mathbf{E}_2$ by the field of the unperturbed problem $\mathbf{E}_1$. This approximation would result in different scaling behaviors for dielectric or metallic ridge inclusions for instance, since both examples correspond to inclusions in the dielectric materials and possess different scaling behaviors for Δε itself. As will be show hereafter, the scaling laws for dielectric or metallic inclusions in the dielectric material or in the metal are all the same. What is wrong with the Born approximation is that local field effects [75] have to be taken into account. This can be achieved on a case–by–case basis, only.

Let us start with slit geometries for instance. Since the $E_x^S$–contribution dominates, the SPP excitation coefficient α is expected to scale as $\alpha \propto k(n_d)^3|n_m|^{-1}\omega\Delta\varepsilon SE_{x,2}|n_m|^{-1}$. It is legitimate to expect that the actual field in the slit is directly proportional to the incident field and thus that $E_{x,2}$ is constant under the scaling. The effective area is proportional to the product of the slit width ($\propto \lambda$) by the skin–depth ($\propto \lambda/n_m$) because line sources located deep into the metal under the interface weakly excite SPP (this can be readily realized by applying the reciprocity theorem [50–51]). Finally, one gets $\alpha \propto |n_m|^{-1}$, and expression that is consistent with Eq. (7).

Let us additionally consider a dielectric ridge. The $E_y^S$–contribution dominates and the SPP excitation coefficient α is expected to scale as $\alpha \propto k(n_d)^3|n_m|^{-1}\omega\Delta\varepsilon SE_{y,2}/\varepsilon_d$. For this simple case, it is easy to realize that $S \propto \lambda^2$, and that $E_{y,2}$ is proportional to the incident field. Again we obtain that $\alpha \propto |n_m|^{-1}$.

In principle, this approach can be generalized to any sub–λ indentation. However this generalization is far from being trivial, because local–field–corrections have to be taken into account. For instance, for metallic ridges, it is difficult to have an intuitive idea of the actual field distribution inside the ridge, so that the scaling behavior of S and $E_{y,2}$ cannot be simply derived without resorting to numerical calculations.

## 4.2 Global approach and numerical tests

Instead of considering a case–by–case approach, it is preferable to adopt a more global approach. The latter relies on the ansatz that the characteristic dimensions of any scatterer all become much larger than the skin depth in the long wavelength regime: as G increases, the scatterer size scales with G but the skin depth of noble metals remains constant according to the Drude model. Thus, as the metal permittivity increases, the scattered–field distribution rapidly tends to the limit case obtained by considering the real metal as perfectly–conducting. One infers that *the scattered field rapidly remains unchanged* as G increases. Thus we obtain

$$\beta(G\lambda) = \beta(\lambda), \tag{22a}$$

for the quasi-CW excitation coefficients. The scaling law is very different for the SPP scattering coefficients. Actually, the SPP contribution to the scattered field vanishes as the permittivity increases. The reason originates from the fact that the SPP–field expands further and further in the dielectric clad as the metal permittivity increases. The SPP–penetration depth into the dielectric medium, $\lambda\chi_{SP}^{-1}\approx\lambda|n_m|/\varepsilon_d$ ($\propto\lambda^2/\varepsilon_d$ according to the Drude model) scales much faster than the purely geometric dilatation G because of the additional metal dispersion. As a result, although the scattered field rapidly reaches its asymptotic value, the SPP–excitation coefficient tends towards zero, with a scaling behaviour given by $\alpha \propto |n_m|^{-1}$

$$\alpha(G\lambda) = [\varepsilon(\lambda)/\varepsilon(G\lambda)]^{1/2} \alpha(\lambda). \tag{22b}$$

We do not repeat here the derivation of this formula that can be established by considering the overlap integral of Eqs. (4a) or (4b) and by replacing the actual field [$E_y(x,y)$ $H_z(x,y)$] by that obtained for the perfectly–conducting case, but the interested reader may refer to [50] for details. The $|n_m|^{-1}$ scaling can be intuitively understood by remarking that it is increasingly difficult to excite a delocalized coherent mode (the SPP mode) from a localized source (the indentation).

Equations (22a) and (22b) are asymptotically valid when $\lambda\rightarrow\infty$. However, in practice they are verified with a good approximation even in the near-infrared (that might depend on the actual indentation geometry and on the incident field). For instance, they well explain why in the slit-doublet experiment of Fig. 1, SPPs are primarily responsible for the fringes at visible and telecommunication wavelengths, whereas they are weakly involved at $\lambda=10$ µm and at longer wavelengths. The SPP−scaling behavior of Eq. (22b) is evidenced by the computational results (Fig. 12) obtained in the visible and near−infrared spectral regions with the frequency−dependent gold permittivities tabulated in [16]. The continuous curves show the modulous of the SPP−excitation coefficient α. The latter is calculated for a normally incident plane−wave illumination and for four geometries, an air−groove, a gold ridge, a silica ridge with a refractive index 1.5 and a metallic post on a dielectric ridge. In agreement with the above analysis, the four curves exhibit the same decrease rate for $\lambda>1$µm, $\alpha \propto |n_m|^{-1}$, which is shown with the black circle marks.

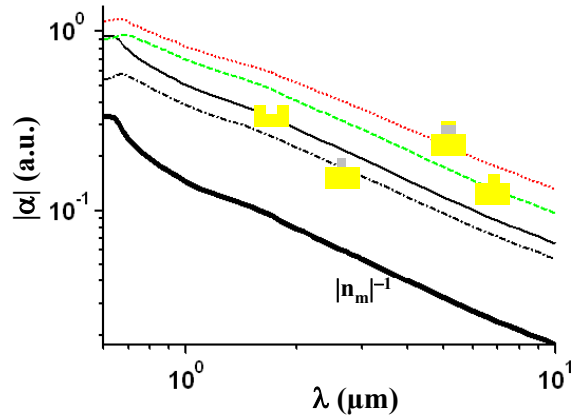

**Fig. 12.** Numerical verification of the scaling laws for the SPP-excitation coefficient α for near-infrared wavelengths and for various geometries. Solid curve: groove with a 0.3λ−width and a 0.15λ−depth, dashed curve: metallic ridge with a 0.45λ−width and a 0.2λ−height, dash-dot curve: dielectric ridge with a 1λ−width and a 0.5λ−height, dotted curve: metallic post on a dielectric ridge (n=1.5) with a 0.3λ−width, a 0.4λ−metallic height and a dielectric (n=1.5) cap 0.25λ−thick. The thick curve represents the inverse square−root of the metal permittivity. The computational results are obtained for gold under normal illumination by scaling the object dimensions with the wavelength. More details can be found in [50].

## 5. SPP and quasi-CW multiscattering processes

Our current understanding of the optical properties of sub−λ metallic surfaces is based on microscopic representations involving multiple−scattering of SPPs [17−34,76], which are assumed to be first generated by some illuminated indentations, then to propagate on the metal surface before being further scattered by nearby indentations. However, this pure SPP picture, although very convenient, is not accurate enough especially at near-infrared wavelengths where quasi-CWs are dominantly excited. In this Section, we present a generalized multiple−scattering formalism [77] that retains all the convenient features of the SPP−model progression picture, but since it combines SPP and quasi-CW waves, it drastically improves the accuracy and enlarges the spectral domain of validity of the classical pure SPP approach.

In order to illustrate the effectiveness and the accuracy of the formalism, we consider the extraordinary optical transmission (EOT) through a periodic hole array drilled in a gold membrane in air. Besides the fact that the EOT is an emblematic phenomenon in plasmonics, this choice has two motivations: First, since efficient electromagnetic numerical tools exist for periodic structures, it allows us to test the accuracy of the present formalism against fully−vectorial computational data. Secondly, since a microscopic SPP couple−mode model [34] has already been elaborated for this phenomenon, one may quantify the net benefit of the present formalism, in comparison to much classical SPP−models.

## 5.1 The extraordinary optical transmission

Ever since its initial description in a famous letter [9], the EOT phenomenon has generated intense interest regarding both the fundamental physics of the transmission mechanism and potential applications. In brief, the EOT effect refers to the observations that light transmission through periodic arrays of sub−λ holes in opaque thin films can be much larger than the combined transmission predicted by standard diffraction theories available for isolated holes in perfectly-conducting thin films [78]. Since the initial publication, the EOT is considered as an emblematic phenomenon of plasmonics, which mix many interesting effects, such as field enhancements on the top and bottom metal-dielectric interfaces of the hole array, light squeezing in small volumes [11], and an intricate Fano-type spectral transmission profiles, see Fig. 13, formed by a sequence of dips and peaks.

Many authors have elaborated on the initial explanation [9] that the excitation of SPPs at either or both metal-dielectric interfaces mediates the EOT effect. Some researchers have promoted a funnelling effect assisted by the excitation of a surface mode on the top or bottom interfaces of the perforated film [79−83]. Some others [84−87] have interpreted the EOT effect in a purely phenomenological fashion, in terms of Fano−type spectral profiles, which result from the interference of two distinct contributions, a resonant channel (the surface modes) and a nonresonant scattering background, and which accounts for the observed spectral profiles peculiar to the detailed geometry of the hole array [88] without explicitly specifying the nature of the resonant channel. In an attempt to fully explain the physical origin of the resonant channel, still other researchers, including two authors of the original contribution, had importantly shown that the transmission enhancement of the hole array should not be compared with that of a single hole in a perfectly-conducting thin film [72]. However, they have been overzealous by completely dismissing the role of SPPs and by putting forth a model based on the generation and the interference of composite diffracted evanescent waves (CDEWs) [72]. A more recent exchange of articles [71,37,89−90], comments [91−92], and responses to comments [93−94,39] between proponents of the SPP and CDEW models has led to a tacit consensus that the CDEW model is incorrect and that both the SPPs and the quasi-CWs are involved in the EOT at visible and near−infrared wavelengths [26], whereas quasi-CWs are the dominant mechanism at longer wavelengths. Our objective is not to add to this difficult discussion here. By considering the EOT phenomenon, we rather aim at illustrating (on a rich toy example) how SPPs and quasi-CWs scatter on nanostructured metal-dielectric interfaces, how we may associate scattering coefficients to these elementary events, and how we may further combine these events to obtain an accurate microscopic description of the macroscopic interface properties.

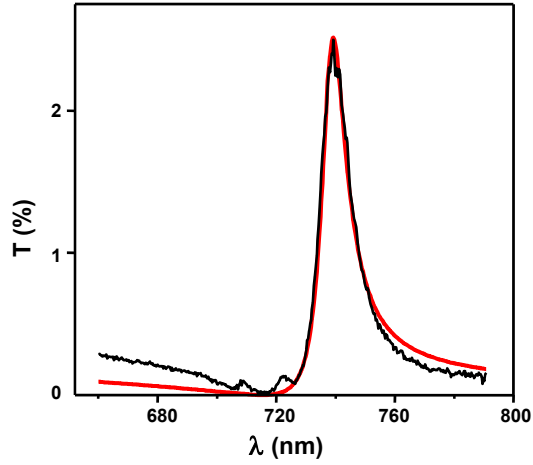

**Fig. 13.** EOT obtained with a gold film of thickness 200 nm, with a grating pitch $a$=700 nm and a hole radius of 70 nm (hole filling fraction 3%). The smooth red curve shows a fitted Fano profile. More details can be found in [85].

## 5.2 SPP–model of the extraordinary optical transmission

Indeed, at a microscopic level, the basic mechanism enabling the EOT is a coherent diffraction by all the individual holes acting as elementary scatterers. However, it is more convenient to consider isolated 2D array of holes (a periodic hole chain with periodicity $a$ in the $y$–direction) perforated in a metal substrate as the elementary scatterers. The related elementary scattering events are shown in Figs. 14a–14d (top row) for classical diffraction geometries (the $y$–component $k_y$ of the in–plane wavevector momentum is null). Upon interaction with the chain, the SPP-modes are partly excited, transmitted, reflected or scattered into the chain mode and into a continuum of outgoing plane waves. The interaction defines four elementary SPP scattering coefficients. Two coefficients, namely the SPP modal reflectance and transmittance coefficients, $\rho_{SP}$ and $\tau_{SP}$, correspond to in–plane elastic scattering, while the two other inelastic ones, $\alpha_{SP}$ and $\beta_{SP}$, transform the SPPs into aperture modes or radiation waves, and vice versa, and allows us to link the local field on the surface to the far–field.

From these elementary–event scattering coefficients, a "pure" SPP coupled-mode model that provides closed–form expressions for the optical transmittance of the periodic array of holes is readily derived. For instance, the reflection coefficient $r_A$ (see the inset in Fig. 15) of the fundamental Bloch mode of the 2D hole array, a very important physical quantity of the EOT phenomenon, can be written [34]

$$r_A = r + \frac{2\alpha_{SP}^2}{u^{-1} - (\rho_{SP} + \tau_{SP})}. \tag{23}$$

In Eq. (23) that holds for normal incidence ($k_x$=0), $u$=exp($ik_{SP}a$) is the phase delay accumulated by the SPP over a grating period and $r$ is the reflection coefficient of the fundamental mode of the hole chain, see Fig. 14b.

The essence of Eq. (23), and in particular of the denominator that results from a geometric summation over all chain contributions, is a multiple scattering process that involves the excitation of SPP modes by the incident field and their further scattering onto the infinite set of periodically-spaced chains. By "pure" SPP–model, we intend to emphasize that the electromagnetic interaction among the chains is only mediated by SPPs of flat interfaces, the quasi-CW being neglected.

The main force of the SPP−model is to make available closed−form expressions for $r_A(k_x)$, and to explain the physical origin of the resonant channel (or surface mode) used in other models. Figure 15 compares the prediction of the pure SPP−model (dashed curve) with fully−vectorial computation results obtained with the well−known Rigorous−Coupled−Wave−Analysis (RCWA) [41,95]. The comparison is performed for three spectral ranges, from the visible ($a$=0.68 μm) to the near-infrared ($a$=2.92 μm). The SPP model quantitatively captures all the salient features of the EOT, and especially the Fano−type spectral profile with the anti−resonance transmission dip and the resonance peak. However, since it does not take into account the quasi-CWs, it is only effective at visible frequencies and becomes highly inaccurate at infrared wavelengths.

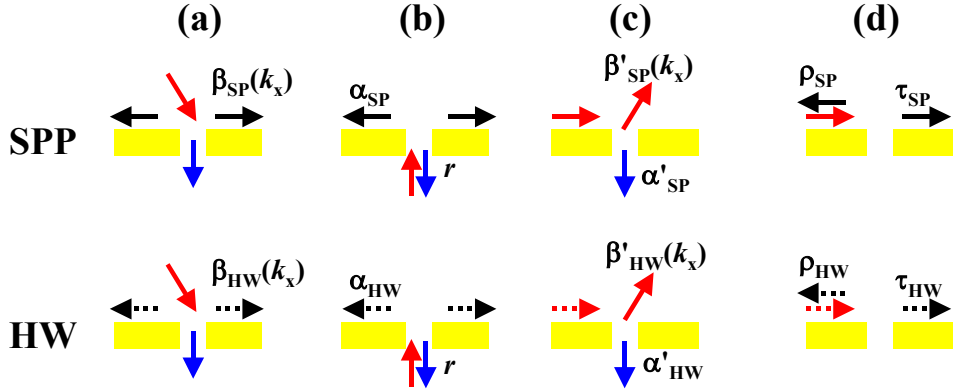

**Fig. 14.** Direct correspondence between SPP (upper row) and HW (lower row) elementary scattering events. (a)−(b) Inelastic scattering coefficients corresponding to SPP or HW excitations by an incident plane wave with a parallel momentum $k_x$ and by the fundamental slit or hole−chain mode. (c) Related coefficients under SPP or HW illumination. According to the reciprocity theorem [50], $\beta'_{SP}(k_x)=\beta_{SP}(k_x)$ and $\alpha'_{SP}=\alpha_{SP}$. (d) Elastic in-plane scattering coefficients. The subscripts 'SP' or 'HW' are related to SPP or HW quantities. More details can be found in [96] and [77].

## 5.3 Generalized hybrid−wave multiple scattering formalism

In this Sub−Section, we present a generalized multiple-scattering formalism that retains all the convenient features of the SPP−model progression picture. However since it combines SPP and quasi-CW waves, it drastically enlarges the accuracy and the spectral domain of validity of the classical SPP approach presented in the previous Section.

The generalized formalism relies on two main ingredients [77]. The first ingredient is related to overall shape of the field scattered by sub−λ indentations on metallic surfaces. This field is always a HW composed of a known mixing ratio of SPP and quasi-CW waves for a given frequency, and the respective contributions are fixed independently of the excitation field and of the exact geometry of the indentation [77]. Physically, this can be understood by considering that, for tiny indentation, the launched SPP and quasi-CW belong to the same scattering process, the latter being impacted by the dielectric property of the metal, rather than by the actual fine structure of the scatterer. This intuitive argument can be consistently justified by considering that, under illumination by an external field $\mathbf{E}^{ext}=E_x^{ext}\mathbf{x}+E_y^{ext}\mathbf{y}$, small nonmagnetic 2D indentations can be represented by effective electric line−dipole sources, either polarized parallel or perpendicular to the interface, $p_x=\alpha_{xx}E_x^{ext}+\alpha_{xy}E_y^{ext}$ and $p_y=\alpha_{yx}E_x^{ext}+\alpha_{yy}E_y^{ext}$, where the alpha coefficients correspond to the polarizability tensor. As shown in Section 3.4, even if the polarizability tensor of sub–λ scatterers strongly depends on the incident illumination and on the indentation geometry, the overall shape (not the amplitude) of the scattered field may be considered as fixed for given energy and metal.

The second important ingredient consists in defining scattering coefficients for HWs. Because quasi-CWs are not normal modes (they are a superposition of modes), they do not obey orthogonality or reciprocity relationships [52]. Thus one may think that it is difficult (or even impossible) to define scattering coefficients for those waves. This difficulty can be overcome for sub–λ indentations by invoking a sort of causality principle [77]. The latter allows us to link the excitation rates and scattering coefficients of HWs and SPPs (Figs. 14a–14d)

$$\beta_{HW}(k_x) \approx \beta_{SP}(k_x),\ \alpha_{HW} \approx \alpha_{SP},\ \beta'_{HW}(k_x) \approx \beta'_{SP}(k_x),\ \alpha'_{HW} \approx \alpha'_{SP},\ \rho_{HW} \approx \rho_{SP},\ \tau_{HW} \approx \alpha_{SP}-1. \quad (24)$$

The causality principle has been recently used to define the cross–conversion between quasi-CWs and SPPs, an inelastic scattering that can be virtually observed whenever a set of sub–λ indentations on a metal surface are illuminated [96].

Equations (24) are remarkably simple and readily relate non–intuitive HW scattering coefficients to much classical SPP coefficients that are routinely calculated with various numerical tools. Additionally, they allow us to preserve the intuitive picture of a microscopic wave progression, and to explicitly analyze the macroscopic properties of metallic surface in terms of a multiple-scattering process. The latter that solely involves local HWs on the flat parts of the surface in between the indentations can be seen as a generalization of the pure SPP–model that involves SPPs only.

The generalized formalism is conceptually very similar to the classical SPP–model, and this conceptual similarity fully reflects in the mathematical analysis. For instance, the new expression for the reflection coefficient $r_A$, which is now [77]

$$r_A = r + \frac{2\alpha_{SP}^2}{(h^{-1}+1)-(\rho_{SP}+\tau_{SP})}, \quad (25)$$

is very similar to that in Eq. (23) obtained with the SPP–model. The only difference is the replacement of $u^{-1}$ by a new term $(h^{-1}+1)$, where $h = \left[\sum_{n>0} H_{z,CW}^+(na,0) + \exp(-ik_{SP}a)-1\right]^{-1}$ encompasses the SPP contribution and an additional contribution corresponding to the interference of all the individual quasi-CWs scattered by the hole chains of the array [77].

The predictions of the generalized formalism are shown with solid curves in Fig. 15. They are much more accurate than those of the SPP–model, and are nearly superimposed with the RCWA data for the three spectral bands, the slight deviation observed at the peak transmission for $a$=2.92 μm and for small wavelengths, being attributed to inevitable numerical inaccuracies or to the single–mode approximation in the hole, rather than to any conceptual problems.

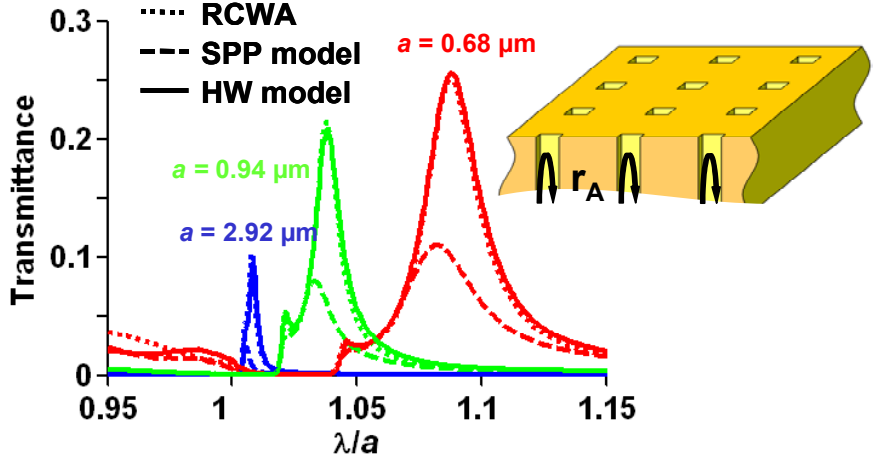

**Fig. 15.** Comparisons between the generalized multiple-scattering formalism predictions (solid curves related to the HW model) and RCWA data (dotted curves) evidence that the formalism is able to predict the EOT phenomenon with a high accuracy over a broad spectral range, from the visible to the near–infrared. The dashed curves show the coupled–mode SPP–model predictions of [34]. The data are obtained for a gold membrane in air perforated by a periodic array of square holes illuminated by a normally incident plane wave. The hole side length $D=0.28a$ (hole filling fraction of 8%) and the membrane thickness is $d=0.21a$, $a$ being the grating pitch equal to $a=0.68\mu m$ (red), $0.94\mu m$ (green) and $2.92\mu m$ (blue). The inset shows the definition of the reflection coefficient $r_A$. More details can be found in [77].

It is important to realize that the generalized HW model does not require additional computations, in comparison with the pure SPP model. Importantly, it relies on the same SPP–scattering coefficients and the additional quasi-CW term, $\sum_{n>0} H^+_{z,CW}(na,0)$, can be calculated analytically, see Eq. (12a). Moreover, its domain of validity is not restricted to periodic surfaces. It is likely to be accurate for non periodic sub–$\lambda$ surfaces, as well.

## *6. Conclusion*

Many optical phenomena related to sub–$\lambda$ metallic surfaces, such as the transmission through arrayed apertures or the oscillatory transmission in slit doublet experiments, which are observed with metallic nanostructures at visible frequencies, can be reproduced at longer wavelengths by scaling the geometrical parameters. At an elementary level, these phenomena are due to the electromagnetic fields that are scattered by the sub–$\lambda$ indentations and that interact with the neighbor indentations. For visible wavelengths, the analysis promotes an interaction mediated by surface–plasmon–polaritons (SPPs) and supplemented at distances of the order of a few wavelengths by an additional scattered near–field, the quasi–cylindrical wave (quasi-CW). At longer wavelength, because they spread far away into the dielectric medium, the delocalized SPPs are marginally excited by the indentations and the quasi-CWs are alone responsible for the existence of the phenomenon [34].

The two–wave model picture represents a helpful microscopic picture for understanding the rich physics of sub–$\lambda$ metallic surfaces. The essence is a multiple scattering process that involves elastic scatterings, CW–to–SPPs cross conversions, and inelastic scattering into radiation modes. For *sub–$\lambda$ indentations* and *good metals*, two important quantities are present at the root of the multiple scattering process: the hybrid–wave (HW), which represents a fixed mix of SPPs and quasi-CWs determined by the 2D Green–function response of the metallic surface, and the elastic and inelastic SPP scattering coefficients, which allows us to keep the important concept of a wave progression on the surface. As such, the HW–formalism represents a generalization of the classical SPP–formalism and quantitatively applies to various geometries and spectral bands, from the visible to the far infrared.

The present analysis has been performed for 2D scatterers such as slits, ridges or sub−λ hole chains. Extension to local scatterers, including various hole shapes, may consistently lead to a similar approach and to a similar physics, relying on a cylindrical radial SPP with a $\rho^{-1/2}\exp(ik_{SP}\rho)$ damping rate and on a quasi−spherical wave with a $\rho^{-2}$ attenuation at long distances [69]. However, nothing is presently known on the scattering of these surface waves and an equivalent of generalized multiple scattering formalism remains to be elaborated.

## *Acknowlegements*

Haitao Liu acknowledges financial supports from the Natural Science Foundation of China (No. 10804057) and from the Cultivation Fund of the Key Scientific and Technical Innovation Project, Ministry of Education of China (No. 708021). Jean Claude Rodier, Lionel Aigouy, Xiao Yang, Jacques Giérak, Eric Bourhis, Christophe Sauvan, Stéphane Collin, Lionel Jacobowiez and Choon How Gan are acknowledged for fruitful discussions and for careful readings of the manuscript.

## *Appendix A*

In this appendix, some details relative to the derivation of the closed−form expression for the quasi-CW fields are given. We start by considering the line−source emission problem shown in Fig. 16, where an infinitely small dipole radiate over an interface. For TM polarization (magnetic field parallel to the *z*−coordinate), the set of Eqs. (8) and (9) can be written

$$\begin{cases} \dfrac{\partial H_z}{\partial y} = -ik\varepsilon E_x + E_x^S\,\delta(x,y) \\ -\dfrac{\partial H_z}{\partial x} = -ik\varepsilon E_y + E_y^S\,\delta(x,y) \quad , \\ \dfrac{\partial E_y}{\partial x} - \dfrac{\partial E_x}{\partial y} = ik H_z + H_z^S\,\delta(x,y) \end{cases} \qquad (A1)$$

where **x**, **y** and **z** are the unitary vectors along the *x,y* and *z* axes, δ(*x,y*) is the 2D Dirac distribution, **E**=$E_x$**x**+$E_y$**y** and **H**=$H_z$**z**.

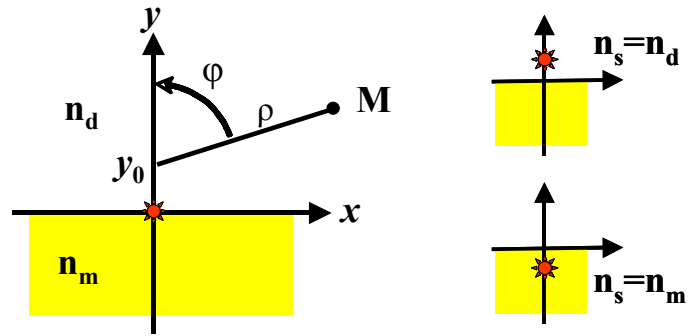

**Fig. 16.** Line source situated at *x*=*y*=0, just above ($n_s$=$n_d$, see inset) or below ($n_s$=$n_m$, see inset) an interface between a dielectric ($n_d^2$=$\varepsilon_d$) and a metallic ($n_m^2$=$\varepsilon_m$) half spaces. The coordinate of point M are *x*=ρsin(φ) and *y*=$y_0$+ρcos(φ), φ∈ [0,π].

Translational invariance in the x–direction allows us to use the inverse Fourier transform, $H_z = \int \hat{H}_z(u,y)\exp(i k u x)\,du$, $\delta(x,y) = \frac{k}{2\pi}\int \delta(y)\exp(i k u x)\,du$ ..., to write the electromagnetic fields with the parallel wavevector $ku$, where $k=2\pi/\lambda$. We readily obtain

$$\begin{cases} \hat{E}_y = \dfrac{u}{\varepsilon}\hat{H}_z + \dfrac{i}{2\pi\varepsilon^S} E_y^S\,\delta(y) \\[4pt] \dfrac{\partial \hat{H}_z}{\partial y} = -i k \varepsilon \hat{E}_x + \dfrac{k}{2\pi} E_x^S\,\delta(y) \\[4pt] \dfrac{\partial \hat{E}_x}{\partial y} = -i k(\varepsilon - u^2)\hat{H}_z - \dfrac{k}{2\pi}\left(H_z^S - \dfrac{u}{\varepsilon_S} E_y^S\right)\delta(y) \end{cases} \qquad (A2)$$

where $\varepsilon_s=\varepsilon_m$ if the source is in the metal, and $\varepsilon_s=\varepsilon_d$ otherwise. Let us define $\chi_d$ and $\chi_m$ by $\chi_d=(\varepsilon_d-u^2)^{1/2}$ and by $\chi_m=(\varepsilon_m-u^2)^{1/2}$, the determination of the square root being such that $\exp(ik\chi_d y)$ and $\exp(-ik\chi_m y)$ represent outgoing waves for $y>0$ and for $y<0$, respectively.

By defining $h$ as $h=ik(u\sin(\varphi)+\chi_d\cos(\varphi))$ for $y>0$ or $h=ik(u\sin(\varphi)-\chi_m\cos(\varphi))$ for $y<0$, we solve the set of Eqs. (A2) and perform the inverse Fourier transform to get a Sommerfeld–type integral

$$[H_z, E_x, E_y] = \int_{-\infty}^{\infty} f(y,u)\exp(\rho h)\,du, \qquad (A3)$$

with

$$f(y,u) = \frac{ik}{2\pi}\frac{\exp(ik\chi_d y_0)}{\dfrac{\chi_d}{\varepsilon_d}+\dfrac{\chi_m}{\varepsilon_m}}\left(H_z^S + \frac{\chi_m}{\varepsilon_m}E_x^S - \frac{u}{\varepsilon_S}E_y^S\right)\left[1, \frac{-\chi_d}{\varepsilon_d}, \frac{u}{\varepsilon_d}\right], \qquad (A4a)$$

for $y>0$, and with

$$f(y,u) = \frac{ik}{2\pi}\frac{\exp(-ik\chi_m y_0)}{\dfrac{\chi_d}{\varepsilon_d}+\dfrac{\chi_m}{\varepsilon_m}}\left(H_z^S - \frac{\chi_d}{\varepsilon_d}E_x^S - \frac{u}{\varepsilon_S}E_y^S\right)\left[1, \frac{\chi_m}{\varepsilon_m}, \frac{u}{\varepsilon_m}\right], \qquad (A4b)$$

otherwise.

The direct calculation of the integral in Eq. (A3) leads, for large $\rho$'s, to the integration of rapidly–oscillating functions. The steepest descent path is the path along which the imaginary part of h remains constant and passes by the saddle point given by $dh/du=0$. By deforming the contour of integration to take the steepest descent path, we reduce at most the impact of the oscillations. The choice of branch cuts that are used to define $\chi_d$ and $\chi_m$ must be compatible with this deformation. We shall not give here the lengthy details of the implementation of the numerical method that allows us to write a stable code calculating completely the integral in Eq. (A3).

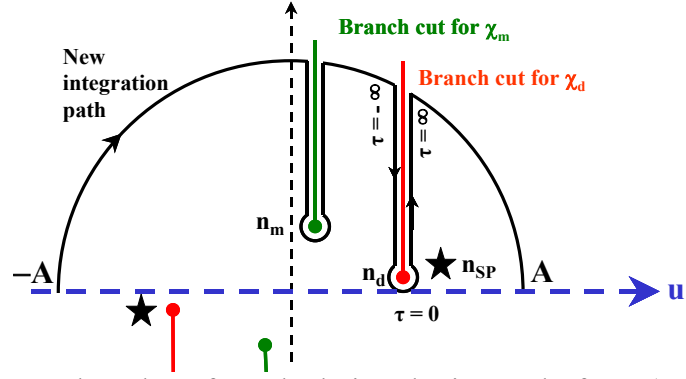

**Fig. 17.** Contour in the complex plane for calculating the integral of Eq. (A3). The black star represents the SPP pole and the branch cuts are shown with green and red vertical lines.

Hereafter, we limit ourselves to an approximate expression of Eq. (A3) in the specific case $\varphi=\pi/2$. Then, $\rho=x>0$, $h=iku$, and $y=y_0$. The branch cuts become half–lines parallel to the imaginary axis and the new integration path is shown in Fig. 17. One may show (but this is not obvious) that, as $A\to\infty$, the contributions of the circle arcs vanish. Cauchy's theorem leads to

$$[H_z, E_x, E_y] = \Phi_{SP}^+ + \Phi_{CW}^+, \tag{A5}$$

where the first term is the SPP contribution of the pole, the second term $\Phi_{CW}^+$ represents the quasi-CW contribution due to the integral over the dielectric branch cuts. The contribution of the metallic branch cut (in green in the figure) is very weak at a certain distance of the origin and it will be neglected hereafter. The residue computation leads to

$$\Phi_{SP}^+(x,y) = ik\frac{n_d^3 n_m^3}{n_d^4 - n_m^4} \exp(ikn_{SP})\left(H_z^S + \frac{\chi_m^{SP}}{\varepsilon_m}E_x^S - \frac{n_{SP}}{\varepsilon_S}E_y^S\right)\begin{cases}\exp(ik\chi_d^{SP}y)\left[1, \frac{-\chi_d^{SP}}{\varepsilon_d}, \frac{n_{SP}}{\varepsilon_d}\right] \\ \exp(-ik\chi_m^{SP}y)\left[1, \frac{\chi_m^{SP}}{\varepsilon_m}, \frac{n_{SP}}{\varepsilon_m}\right]\end{cases}, \tag{A6}$$

where the top (resp. bottom) expression holds for $y>0$ (resp. $y<0$). Equation (A6) is equivalent to Eq. (11).

The calculation of the dielectric branch cut deserves more attention. We first parameterize the branch–cut path using a parametric equation $u=n_d+i\tau^2$, where $\tau$ is real, varies from $-\infty$ to $\infty$ and is null at the branch point $u=n_d$. The new pole $\tau_{SP}$ can be written $\tau_{SP}=\exp(-i\pi/4)(n_{SP}-n_d)^{1/2}$ and $\chi_d=-\tau(\tau^2-2in_d)^{1/2}$. The classical steepest descent method consists in expanding the integrand in Eq. (A3) in the vicinity of $\tau=0$ and in keeping the constant term, only. In the limit $x\to\infty$, the classical steepest descent method provides a rigorous asymptotic expansion with a $\rho^{-3/2}$ damping rate, but for finite $x$, it is better using a first–order expansion, $(\tau-\tau_{SP})f(\tau)\approx(\tau-\tau_{SP})f(0)-\tau_{SP}\tau(df/d\tau)_{\tau=0}$, which accurately handles the close proximity of the pole $\tau_{SP}$. This approach is refered to as a modified steepest descent method in [55,70,97]. The quasi-CW contribution is then approximately given by

$$\Phi_{CW}^+(x,y) \approx \exp(ikn_d x)\int_{-\infty}^{\infty} 2i\tau[f(0)-\tau_{SP}\tau/(\tau-\tau_{SP})(df/d\tau)_{\tau=0}]\exp(-k\rho\tau^2)\,d\tau. \text{ By setting}$$

$$W(t) = -2\sqrt{\frac{-it}{\pi}} \int_{-\infty}^{\infty} z^2 \frac{\exp(-z^2)}{z - \sqrt{-it}} dz, \tag{A7}$$

we obtain

$$\Phi_{CW}^{+}(x,y) \approx i\sqrt{\pi}\exp(ikn_d x)\left(\frac{df}{d\tau}\right)_{\tau=0} \frac{W(\gamma k x)}{kx\sqrt{kx}}, \tag{A8}$$

with $\gamma = n_{SP} - n_d$. Other lengthy calculations lead to

$$\left(\frac{df}{d\tau}\right)_{\tau=0} = \frac{k}{2\pi}\sqrt{2in_d}\frac{\varepsilon_m^2}{\varepsilon_d(\varepsilon_m-\varepsilon_d)}\left(H_z^S + \frac{\chi_m^0}{\varepsilon_m}E_x^S - \frac{n_d}{\varepsilon^S}E_y^S\right)\left[1 - iky\frac{\varepsilon_d\chi_m^0}{\varepsilon_m}, \frac{\chi_m^0}{\varepsilon_m}, \frac{1}{n_d}\left(1 - iky\frac{\varepsilon_d\chi_m^0}{\varepsilon_m}\right)\right]$$
, (A9a)

for $y>0$ and

$$\left(\frac{df}{d\tau}\right)_{\tau=0} = \frac{k}{2\pi}\sqrt{2in_d}\frac{\varepsilon_m^2}{\varepsilon_d(\varepsilon_m-\varepsilon_d)}\left(H_z^S + \frac{\chi_m^0}{\varepsilon_m}E_x^S - \frac{n_d}{\varepsilon^S}E_y^S\right)\left[1, \frac{\chi_m^0}{\varepsilon_m}, \frac{n_d}{\varepsilon_m}\right], \tag{A9b}$$

for $y<0$. In the previous equations, $\chi_m^0 = \sqrt{\varepsilon_m - \varepsilon_d}$. Finally, we obtain

$$\Phi_{CW}^{+}(x,y) \approx k\sqrt{\frac{in_d}{2\pi}}\frac{\varepsilon_m^2}{\varepsilon_d(\varepsilon_m-\varepsilon_d)}\frac{W(\gamma kx)}{kx\sqrt{kx}}\left(H_z^S + \frac{\chi_m^0}{\varepsilon_m}E_x^S - \frac{n_d}{\varepsilon^S}E_y^S\right)\Psi_0^{+}, \tag{A10}$$

with $\Psi_0^{+}$ defined in equations (14a) and (14b).

The function $W(t)$ slowly varies with $t$ and is equal to $2it - 2t\sqrt{-i\pi t}\exp(it)[\text{Erf}(i\sqrt{-it}) + \text{sign}(\text{Im}(\sqrt{-it}))]$, with $\text{Erf}(t) = \frac{2}{\sqrt{\pi}}\int_0^t \exp(-z^2)dz$ being the error function. Some other properties are $W(t) \to 1$ as $t \to \infty$ and $W(t) \propto 2it$ for $t \to 0$.

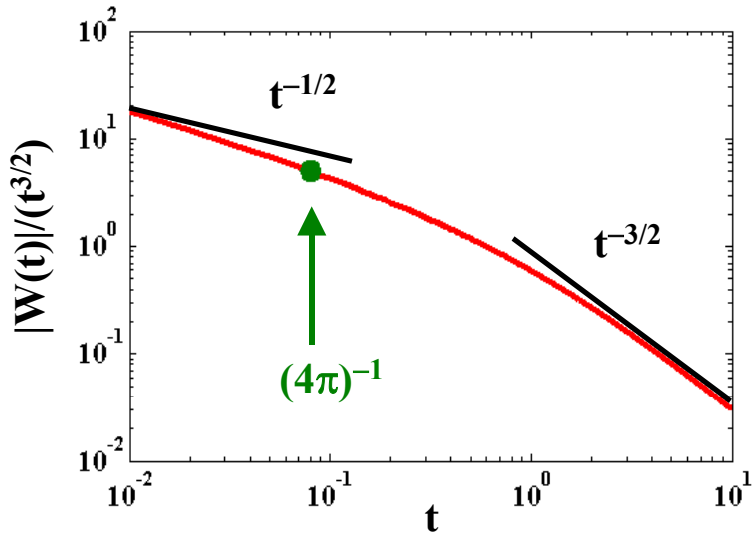

**Fig. 18.** Plot of $\frac{|W(t)|}{t\sqrt{t}}$ as a function of t real. The transition from the initial $\frac{1}{\sqrt{t}}$ damping regime to the asymptotic $\frac{1}{t\sqrt{t}}$ damping regime takes place near $4\pi t=1$. W(t) can be calculated with Matlab as W=@(t) -2*sqrt(-i*t/pi)*quadgk(@(z) z.^2.*exp(-z.^2)./(z-sqrt(-i*t)),-inf,inf) [98].